%% file: revisexx.tex
\documentclass[sn-mathphys-num]{sn-jnl}
\usepackage{graphicx} %
\usepackage{multirow}%
\usepackage{amsmath,amssymb,amsfonts}%
\usepackage{amsthm}%
\usepackage{mathrsfs}%
\usepackage[title]{appendix}%
\usepackage{xcolor}%
\usepackage{textcomp}%
\usepackage{manyfoot}%
\usepackage{booktabs}%
\usepackage{algorithm}%
\usepackage{algorithmicx}%
\usepackage{algpseudocode}%
\usepackage{listings}%

\newcommand\eref[1]{Eq.~(\ref{#1})}
\newcommand\Eref[1]{Equation~(\ref{#1})}
\newcommand\fref[1]{Fig.~\ref{#1}}

\newcommand\sref[1]{Sect.~\ref{#1}}
\newcommand\Sref[1]{Section~\ref{#1}}
\DeclareMathOperator{\sgn}{sign}
\DeclareMathOperator{\prob}{Prob.}

\usepackage[normalem]{ulem}

\usepackage{etoolbox}
\makeatletter
\patchcmd{\maketitle}{Contributing authors:}{Emails:}{}{}
\patchcmd{\@maketitle}{Contributing authors:}{Emails:}{}{}
\makeatother

\begin{document}

\title{Hysteresis  in magnets}

\author[1]{\fnm{Deepak} \sur{Dhar}}
\email{deepakdhar1951@gmail.com}

\author[2]{\fnm{Sanjib} \sur{Sabhapandit}}
\email{sanjib@rri.res.in}

\affil[1]{\orgdiv{International Centre for Theoretical Sciences}, \orgname{Tata Institute of Fundamental Research}, \orgaddress{\city{Bangalore}, \postcode{560089}, \country{India}}}

\affil[2]{\orgname{Raman Research Institute}, \orgaddress{\city{Bangalore}, \postcode{560080}, \country{India}}}

\abstract{We provide an overview of studies of hysteresis in models of magnets.  We discuss the shape of the hysteresis loop, dynamical symmetry breaking, and the dependence of the area of the loop on the amplitude and frequency of the driving field. We also discuss Barkhausen noise in  the hysteresis loops, where the wide distribution of sizes of  magnetization jumps may be modeled by   the random field Ising model. We discuss the distribution of  sizes of these jumps in the random field Ising model on the Bethe lattice. 
}

\maketitle

\section{Introduction}

In this article, we  discuss hysteresis  in magnets and Barkhausen noise.
There are many existing reviews of these in the literature.  Our aim here is not to summarize the work done since the last review, but to provide a pedagogical self-contained overview. For a detailed discussion of the literature, the readers may consult earlier reviews and more recent works that have cited them~\cite{review1, review2, review3, lyuksyutov, sides1, sides2a, sides2b, sides2c, Colaiori08,mayergoyz2006science}. 

This review is organized as follows: In~\sref{s:early-work}, we discuss briefly some early work of Landau--Lifshitz--Gilbert and Preisach.  In~\sref{s:scaling-laws}, we discuss the scaling laws for hysteresis loop area, and start with some simple examples where such scaling laws can be inferred directly.  In~\sref{s:Ising-like}, we discuss the scaling laws for the hysteresis in a system of Ising spins. In~\sref{s:continuous-spin}, we discuss the continuum spins case. In~\sref{s:dpt}, we discuss the dynamical phase transition in both Ising and continuous spin systems. 
In~\sref{s:RFIM}, we consider hysteresis in strongly disordered Ising systems, first in the mean field theory, and then on the Bethe lattice.  In~\sref{s:avalanche}, we discuss the distribution of avalanche sizes in the hysteresis in the random field Ising model on the Bethe lattice.  \Sref{s:conclusion} contains some concluding remarks.

\section{Early work}
\label{s:early-work}

The phenomenon of hysteresis has been studied experimentally for a long time. There were also theoretical explanations that describe qualitatively the observed 
behavior. There have been many such attempts in the past, with varying degrees of success.  They do capture much of the phenomenology, but are not fully satisfactory from a theoretical point of view, as they are equivalent to some closure approximation for the hierarchy of equations for the $n$-point correlation functions (i.e., the Bogoliubov--Born--Green--Kirkwood--Yvon hierarchy). Let us mention two here: The Landau--Lifshitz--Gilbert equations and the Preisach model.

\subsection{The Landau--Lifshitz--Gilbert equation}

The Landau--Lifshitz--Gilbert equation \cite{landau, lakshmanan}, for the evolution of a spin $\vec{S}(t)$ is given by 
\begin{equation}
\frac{d}{dt} \vec{S}(t) = - \Gamma \left[ \vec{S}(t) \times \vec{h}_\mathrm{eff}(t)\right] + \lambda \Gamma \left[ \vec{S}(t) \times \frac{d}{dt} \vec{S}(t) \right],
\label{eq:LLG}
\end{equation}
Here $\vec{h}_\mathrm{eff}$ is the effective field felt by the spin, and describes the precession of the spin. This first term may be derived from the microscopic quantum mechanical  Hamiltonian (or its classical approximation).  The second term was introduced by Gilbert to describe the damping of the spin that tends to make it align parallel to the effective field.   However,  to make the equation tractable, one usually makes the closure approximation that $h_\mathrm{eff}(t) $ is proportional to the magnetization $M(t)$. This equation describes well relaxation in ferromagnetic materials and also is of interest as one can get some exact solutions, e.g.,  for a single soliton on a linear chain in one, and also in higher dimensions.  See~\cite{lakshmanan} for details.

\subsection{The Preisach model}

In the Preisach model \cite{preisach}, one treats the magnetic material as a disordered system, made up of many independent domains. The domains have a distribution of sizes.  The response to the external field $h(t)$ on any domain is approximated by a rectangular hysteresis loop. Let us denote the magnetization of the $i$-th domain during the time when the field is being increased from $-\infty$ to $h(t)$  by $M_i^\mathrm{incr}(h(t))$.  In this approximation, it is  given by
\begin{align}
M_i^\mathrm{incr}(h(t)) =&  - M_{0i}, \quad\text{for}\quad h< h_{0i};\\
       =&  + M_{0i}, \quad\text{for}\quad h> h_{0i},
\end{align}
where $M_{0i}$ and $h_{0i}$ are taken to be independent identically distributed  random variables with given distributions.  These take into account  approximately the  effect of differences between the sizes of domains and the exchange interactions between domains. During the part of the cycle when the field is being decreased, the magnetization is given by $M_i^\mathrm{decr} (h) = - M_i^\mathrm{incr}(-h)$.  Knowing the probability distribution, one can calculate the response of the magnet to any time-dependent field $h(t)$.  For details, see~\cite{vistitin,mayergoyz}.

\section{Scaling of the area of the hysteresis loop with frequency and amplitude of the applied field}
\label{s:scaling-laws}

 In applications, such as a magnetic material subjected to a periodically changing magnetic field, the energy that is dissipated per cycle is an important consideration.  This is equal to the area of the hysteresis loop in the $h$-$M$ (field magnetization) plane, and its variation with the peak magnetic flux density $B_0$ and frequency $\omega$ of the imposed external field has been a matter of intense investigation for over 130 years~\cite{steinmetz}. 
 
Steinmetz empirically proposed that the power dissipation varies as $ B_0^{\alpha} \omega^{\beta} $. This included the effect of magnetic work (area of the hysteresis loop) and eddy current losses. In practice, the fitted values of the so-called  `Steinmetz coefficients' $\alpha$ and $\beta$ were generally found to be temperature-dependent even for a given material. In the following, we  study the variation of only the area of the hysteresis loop. The eddy current losses are more complicated and should be discussed separately. The dependence of the loop area on the temperature is also nontrivial.  Clearly, at large temperatures $T$,  much above the Curie temperature, the response, and hence also area, is proportional to $1/T$, but for lower temperatures, the response can be much more complicated~\cite{acharyya2}. For low temperatures, in continuous spin models, one can expect the predictions of the harmonic approximation (independent spin waves) to be good,  and  theoretical calculation  for the  variation of the barrier-crossing times  for the time-dependent Hamiltonian  on the temperature   to be reliable (see  \sref{s:continuous-spin}). For Ising-type models, similarly, the rates of relaxation is governed by the activation barriers.  We do not discuss the temperature dependence of the loop area here. 

\begin{figure}
\centering
\includegraphics[width =\linewidth]{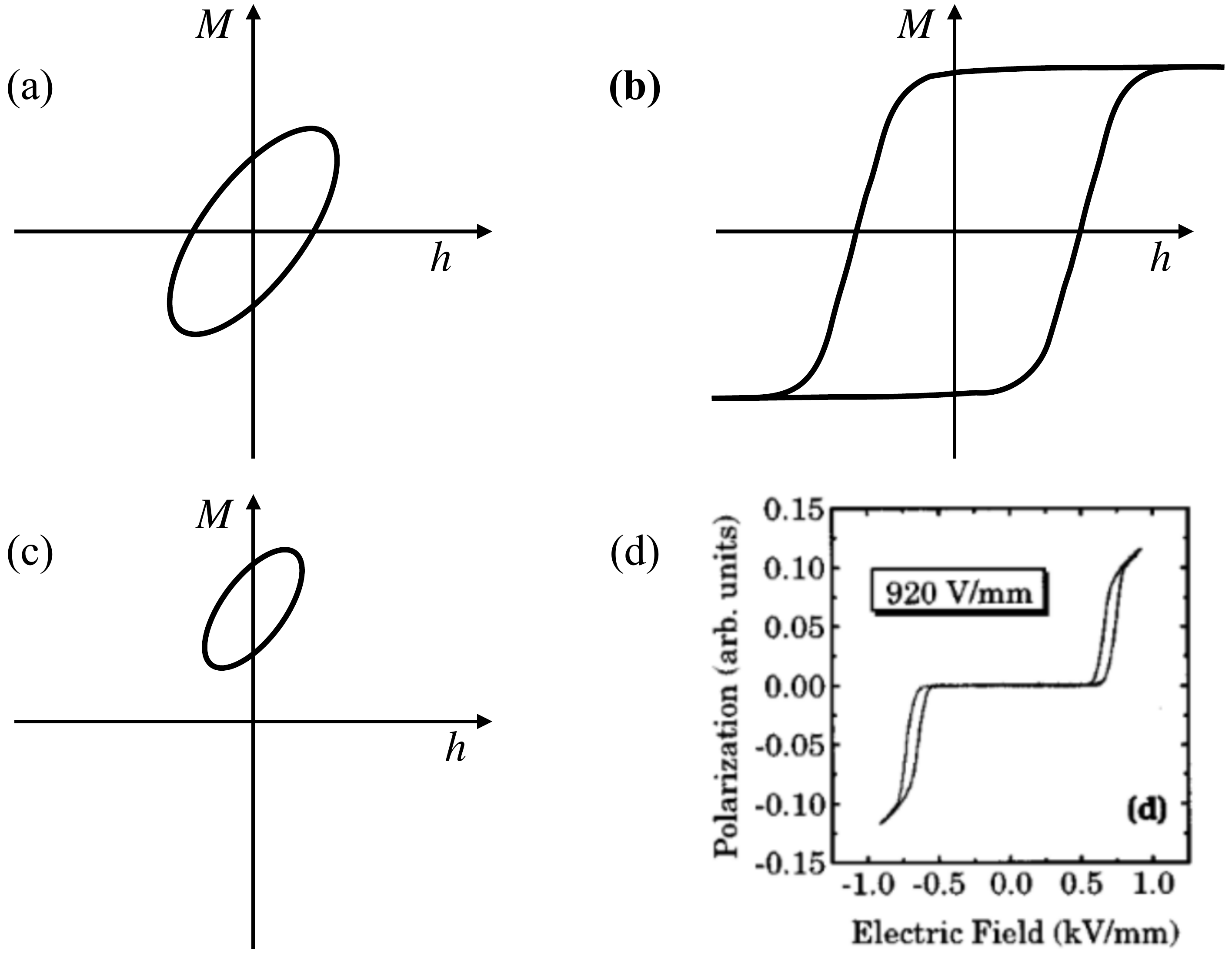}
\caption{Different types of hysteresis loops.  (a) Paramagnetic, (b) ferromagnetic symmetric, (c) ferromagnetic asymmetric,  (d) `wasp-waisted'  loop in antiferroelectric mixed crystal of betaine phosphate-arsenate ($BP_{0.9}As_{0.1}$) as a function of the electric field. 
Reprinted figure (d) with permission from~\cite{kim}~[\href{https://doi.org/10.1103/PhysRevB.55.R11933}{Y.-H. Kim and J.-J Kim, Phys. Rev. B {\bf 55}, R11933  (1997)}]. Copyright (1997) by the
American Physical Society.
}
\label{fig:fig1}
\end{figure}

From a theoretical point of view, one would expect such scaling laws to be material or temperature independent, and show some degree of universality.  But this universality would be expected only in some asymptotic power laws, and effective exponents would be expected to show much more complicated behavior involving non-universal corrections to scaling.  The asymptotic power laws can thus be expected only if we first take the thermodynamic limit, and then  only in the limits of $\omega$ or $B  \rightarrow 0$ or $ \infty$.\footnote{Near a dynamical phase transition, we could expect power-laws in {\em deviations from the transition point values.}}  It is also theoretically preferable to work with the strength of the imposed external field $h_0$, rather than the field $B$, in which the sample magnetization also contributes. For our  discussion here, we take  the exponents $\alpha $ and $\beta$ as defined by  the condition that  the area of the hysteresis loop $A$ varies as $h_0^{\alpha} \omega^{\beta}$, in the appropriate limit. Similar considerations apply to hysteresis in a ferroelectric system as a function of imposed electric field~(\fref{fig:fig1}).

To fix notation, let us first define the system under consideration precisely. We consider a macroscopic magnetic sample in contact with a heat bath at a temperature $T$, or equivalently at inverse temperature $\beta=1/(k_\mathrm{B} T)$.\footnote{With a slight abuse of notation, 
the symbol $\beta$ is also used in this review to denote one of the Steinmetz coefficients introduced earlier. Its intended meaning should be clear from the context.} It is subjected to a sinusoidally varying external magnetic field  $h(t) = h_0 \sin \omega t$.  If we let the system evolve for a long time, the system is expected to settle into a  {\it periodic} steady state of the system that varies with the same frequency, and the time-dependent magnetization per unit volume 
$M(t)$ is a periodic function of time with period $2 \pi/\omega$.  The plot of $M(t)$ versus $h(t)$ then is a closed curve called the steady state hysteresis loop~(\fref{fig:fig1}). We study the behavior of this response as a function of the control parameters of the problem. The dynamics of the spin system is assumed to be a Markovian evolution, in which a single spin can change its state at a rate that satisfies the detailed balance condition.  The latter condition ensures that in the limit of  $\omega \rightarrow 0$, the steady state is in thermal equilibrium at all times.

\subsection{ Some simple cases}

The limiting behavior of the hysteresis loop area is easily determined in some simple cases. We discuss these first. The limit of $h_0$ very large is not easily realized, but in simple models like the Ising model, the magnetization has a maximum possible value unity. Since the response of the system has  a finite delay time,  the area scales as $h_0$. In theoretical models like the $\phi^4$ model,  where $\phi$ can take arbitrarily large values, and the free energy density as a function of the magnetization is of the form $F(M, h)  \sim a M ^2 + b M^4 - h M$, it is easy to see that 
for large $h_0$, the time for the magnetization to switch  still remains finite,  of $O(1)$, and the coercive field increases as  $h_0$.  As the maximum magnetization  in the Ising and $\phi^4$ cases    increases as $h_0^0$ and $h_0^{1/3}$, respectively, the area exponent $\alpha$   takes the values  $1$ and $4/3$, respectively, in these cases.

\begin{figure}
    \centering
    \includegraphics[width=\linewidth]{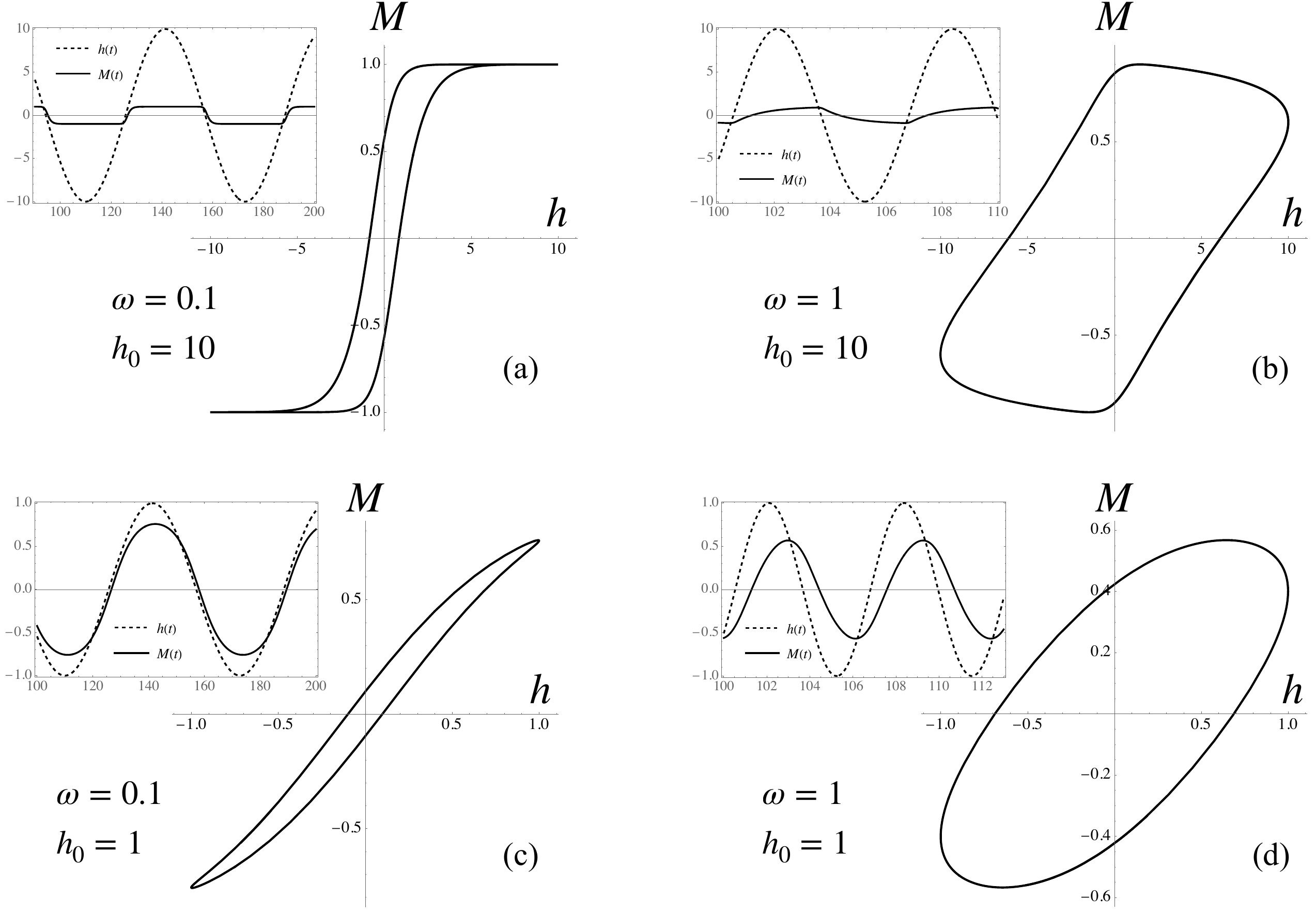}
    \caption{Plot of the hysteresis for a single spin driven by the oscillating magnetic field $h(t) = h_0\sin(\omega t)$, where the corresponding time-dependent magnetization is given by \eref{eq:M2-1spin}. The time dependence of $h(t)$ and $M(t)$ is shown in the corresponding insets. (a) low frequency--high field amplitude, (b)  high frequency--high field amplitude, (c) low frequency--low field amplitude, and (d) high frequency--low field amplitude.}
    \label{fig:hys-1spin}
\end{figure}

To understand the change in the shape of the loops with $h_0$ and $\omega$, it is instructive to look at the dynamics of a single Ising spin $s=\pm 1$ under an oscillating magnetic field $h(t)=h_0 \sin(\omega t)$. The system is described by  a simple Hamiltonian
\begin{math}
H = -h(t) s. 
\end{math} 
We assume that the system
evolves according to the Glauber dynamics~\citep{Glauber1963}, where the spin
flips from $s$ to $-s$ at a rate
\begin{math} 
W(s\rightarrow -s)=(\Gamma/2)\bigl(1-s \tanh\bigl[\beta\,h(t)\bigr]\bigr). 
\end{math} 
Given the Glauber's spin flip rate, the average magnetization $M(t)\equiv\langle  s(t)\rangle$ obeys the linear
differential equation~\cite{Glauber1963}  
\begin{equation}
\frac{dM}{d t} + \Gamma M =\Gamma \tanh\bigl[\beta\, h(t)\bigr].
\label{eq:mt1}
\end{equation}
This equation is easily integrated using the integrating factor $e^{\Gamma t}$ to obtain an explicit solution.  Since $-1\le \tanh(x)\le 1$, it is easy to see that late times, taking the initial time of preparation $t_0 \to -\infty$,   the solution  converges to 
\begin{equation}
M(t)=\Gamma \int_0^\infty\, e^{-\Gamma t_2}\, \tanh(\beta h(t-t_2))\, dt_2\, .
\label{eq:M2-1spin}
\end{equation}
This equation is valid for any function $h(t)$. For a periodic driving field $h(t)$ with a period $2\pi /\omega$, it is evident from~ \eref{eq:M2-1spin} that $M(t)$ is also a periodic function of $t$ with the same time period $2\pi /\omega$. 
The time-dependent magnetization in the time-periodic steady state is demonstrated in~\fref{fig:hys-1spin} for the driving field $h(t)=h_0\, \sin(\omega t)$.

More generally, the linear response theory gives the correct leading behavior in the case of hysteresis in paramagnets, or in systems at high temperature.  The evolution equation may be
linearized as 
\begin{equation}
\frac{d}{dt} M(t) = - a M(t) +  h_0 \exp(i \omega t)
\end{equation}
where it is convenient to use  the  complex time dependence  $\exp( i \omega t)$. This is a linear equation, and its real part gives  $M(t) =    h_0  \,\mathrm{Re}\left[  \exp(i \omega t)/( a + i \omega) \right]$. This implies that the scaling for the area of the hysteresis loop is  $A = - \oint\,M\,dh \sim h_0^2/ \omega$ for large $\omega$.  

When there is spontaneous magnetization, the above equation has to be modified.  In the simplest approach, we  write down an (approximate) evolution equation for the bulk magnetization: 
\begin{equation}
\frac{d}{dt} M(t) = a M - b M^3 + h_0 \sin  \omega t. 
\label{eq:jung}
\end{equation}
If we linearize this equation about any one of the two possible equilibrium values in zero field $\bar{M} = \pm  \sqrt{a/b}$, the deviation from the mean $\delta M(t) = M(t) - \bar{M}$ varies as $ h_0/\omega$, and the loop area still scales as $ h_0^2/\omega$. 

In~\fref{fig:hys-1spin}, we illustrate the hysteresis loops for the single-spin model described by~\eref{eq:mt1}, by plotting $M(t)$ in~\eref{eq:M2-1spin}  vs $h(t)=h_0 \sin(\omega t)$ parametrically as a function of $t$, for different $\omega$ and $h_0$. It is evident from  \fref{fig:hys-1spin} that the area of the hysteresis loop decreases  to zero as $\omega\to 0$. 
The area of the hysteresis loop in the time-periodic steady state is given by
\begin{equation}
A =- \oint\,M\,dh =-\int_{0}^{2\pi/\omega}M(t)\, \frac{dh(t)}{dt}\, dt.
    \label{eq:A1}
\end{equation}
Substituting the expression of $M(t)$ from~\eref{eq:M2-1spin} and $h(t)=h_0 \sin(\omega t)$ in the above equation, after integration by parts and simplification, we get
\begin{equation}
A=\frac{4 \Gamma h_0 \omega}{ \Gamma^2+\omega^2}\int_0^{\pi/2}\,
d\theta\,\sin\theta\,\tanh(\beta h_0\sin\theta)
\label{i1}
\end{equation}
The integral in \eref{i1} is hard to perform in closed form
 to obtain an explicit expression for $A$. However, using the properties   
$\tanh(\beta h_0\sin\theta)\approx\beta h_0\sin\theta$ for $\beta h_0 \ll 1$ and $\tanh(\beta h_0\sin\theta)\approx 1$ for $\beta h_0 \gg 1$, we can easily perform the integral in these two limits. Consequently, the area in these two limits is given by 
\begin{equation}
A\approx\begin{cases}
\displaystyle
   \frac{\pi \Gamma\beta h_0^2 \omega}{ (\Gamma^2+\omega^2)}  & \text{for~} \beta h_0 \ll 1,\\[5mm] 
   \displaystyle
   \frac{4 \Gamma h_0 \omega}{ (\Gamma^2+\omega^2)} 
   &\text{for~} \beta h_0 \gg 1.
\end{cases}
\end{equation}

\section{ Hysteresis  in Ising-like models}
\label{s:Ising-like}

Now, we discuss the more interesting case of small frequencies in models where the magnetization is treated as a scalar field. In this case, there has been some confusion in the literature. There are two different situations that are discussed, but not clearly distinguished. In the first case, it was realized that the qualitative behavior of the problem is independent of temperature, and so, one may simplify the problem and study the dynamics at zero temperature. Then, if we now consider the limit $\omega \rightarrow 0$, {\em the area of the hysteresis loop tends to a non-zero value.} Note that for nearly rectangular ferromagnetic loops, the amplitude of the field has to be sufficiently large (larger than than the coercive field value), for the area of the loop to be non-zero.

In the second case, one studies the problem at non-zero temperatures, and then in the same limit, if the rate of change of the field is slow enough,  the system tends to thermodynamic equilibrium at each value of the field.  Then, as the field changes sign from negative to positive, the magnetization shows a discontinuous jump, from $ - M_0$ to $+M_0$, but the area of the hysteresis loop tends to zero. {\em Clearly the limits $T \rightarrow 0$ and $\omega \rightarrow 0$ cannot be interchanged.} 

\subsection{ The  bistable oscillator driven by  an oscillating field}

Let us discuss the first case first.  This was discussed by \citet{jung}. They argued that magnetic systems show bistability, and considered the simplest evolution for single variable $ M(t)$ given by~\eref{eq:jung}.

This equation can be solved exactly, and the solution can be expressed in terms of Airy functions. One finds that if the amplitude of the field $h_0$ is small, then $M(t)$ stays near one of the  minima, and only for $ h_0$ greater than a value $h_m$  does one get an inversion symmetric hysteresis loop.  In this case, they found that  the area of the loop $A( h_0,\omega)$ varies as 
\begin{equation}
A(h_0,\omega) = K_1 + K_2 \left[ ( h_0^2 - h_m^2) \omega^2 \right]^{1/3}, \quad\text{for}\quad h_0^2 > h_m^2,
\end{equation}
where $K_1$ and $K_2$ are some constants depending on $a$ and $b$ in~\eref{eq:jung}. This is sometimes quoted as $\omega^{2/3}$-dependence of the hysteresis area, but note that the area goes to a non-zero constant as $\omega$ tends to zero, and so, the leading behavior corresponds to  $\beta = 0$. 

If $h_0^2 < h_m^2$, then the hysteresis loop is not unique and does not have inversion symmetry. This is a symmetry-breaking non-equilibrium phase transition, discussed in more detail in~\sref{s:dpt}. 

The analysis of Jung et al., where the hysteresis response  can be described by the evolution \eref{eq:jung}, is an ordinary differential equation in a  single  variable,  which has been generalized by Goldzstein et al.,~\cite{goldzstein} to the case where the equation of evolution of the single macrovariable  $x(t)$ is of the form 
\begin{equation}
\frac{d}{dt} x(t) = - F(x) + h_0 \sin \omega t,
\end{equation}
where $F(x) $ is a continuous {\em monotonically increasing} function of $x$. They show that there is a unique periodic solution to which all other solutions tend to in the large time limit, and there is no dynamical symmetry breaking in this case. They show that in the case $F'(0) >0$, the area of the hysteresis loop is proportional to $\omega$ in the $\omega \rightarrow 0$ limit. They also discuss the case when $F(x) $ varies as $x |x|^a$, then the exponents $\alpha$ and $\beta$ depend continuously on $a$. Note that in the case of the $n$-vector  model in an oscillating field, the evolution equations are not reducible to a single variable equation without noise, and the above arguments do not apply.

\subsection{Hysteresis governed by droplet-nucleation dynamics in Ising-like models}

\begin{figure}
\centering
\includegraphics[width=.6\linewidth]{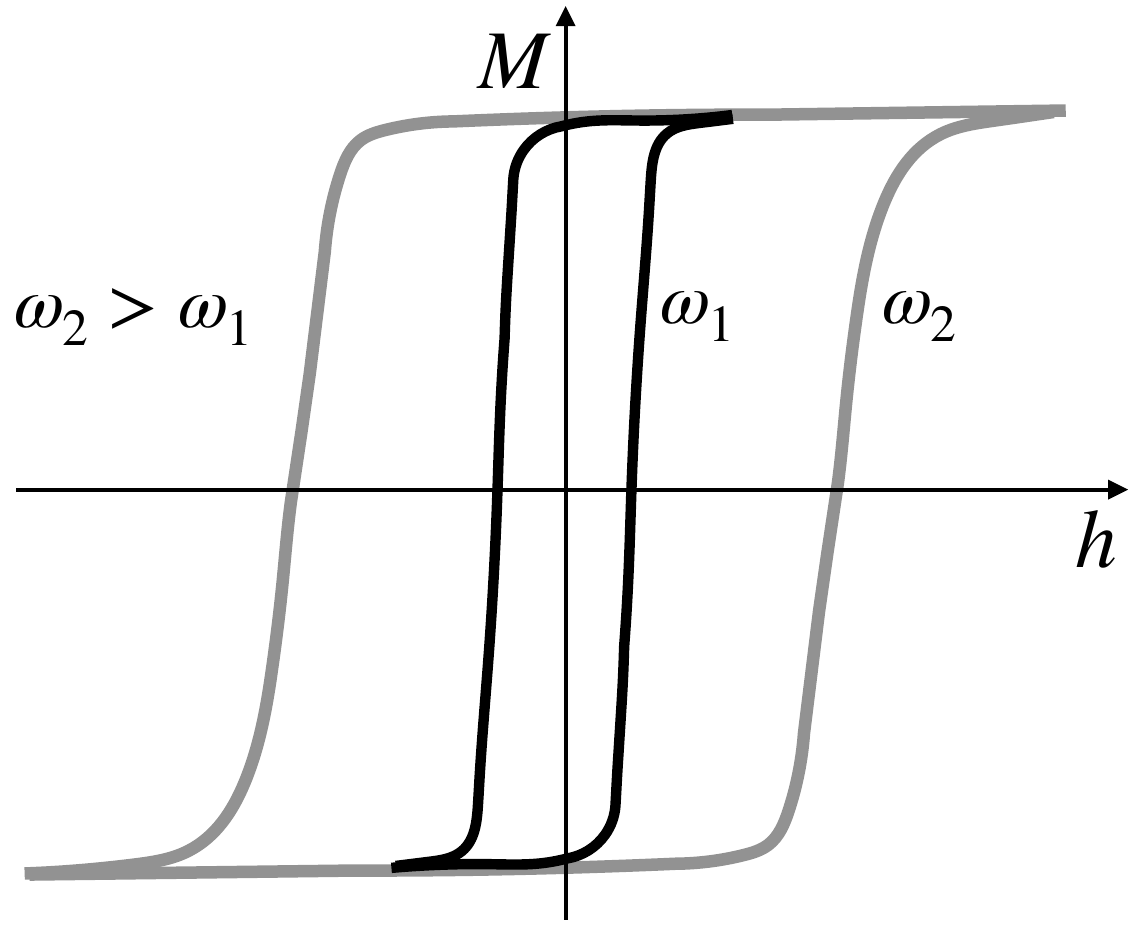}
\caption{Qualitative behavior of the hysteresis loop in the $d$-dimensional Ising model at very small frequencies.  The loops are nearly rectangular, and the area decreases with decreasing frequency.}
\label{fig:loops2}
\end{figure}

Now, let us consider the case of non-zero temperatures, where in the small $\omega$ limit, the area would be expected to go to zero (see~\fref{fig:loops2}).

If we keep $h_0$ constant, and decrease $\omega$,  the change in magnetization occurs for small fields $\ll h_0$.  In this limit, the  part of the hysteresis loop where the magnetization changes significantly is nearly the same, i.e., when $|h(t)|$  is  small, if we replace the sinusoidal time dependence by a linear dependence.  We may then also allow the maximum magnitude of the magnetic field to be infinite, without affecting the loop shape and take  
\begin{equation}
h(t) = h_0\, \omega t, \quad\text{for}\quad -\infty \le t \le + \infty.
\label{eq:field}
\end{equation}
This then implies that  the area $A( h_0,\omega)$ is a function of the single variable $ h_0 \omega$, in the $\omega\to 0$ limit, keeping $h_0\omega$ constant:
\begin{equation}
A( h_0, \omega) = f( h_0 \omega)  + [\text{terms of } o(\omega)].
\end{equation}

To be specific, let us consider an Ising model with nearest-neighbor ferromagnetic couplings on a $d$-dimensional hypercubical lattice.  The time evolution is Markovian, with single-spin flip Glauber dynamics. Any spin slips from its current value $S_i$ to $-S_i$ at a rate $ (\Gamma/2) [ 1 - \tanh( \beta \Delta E_i/2)]$,  where $\Delta E_i$ 
is the change in energy due to the flip.  The system is subjected to a time-dependent external field, given by Eq. (\ref{eq:field}).

First, we discuss the case of symmetrical hysteresis loops. For large negative values of the field, all the spins are down.  As the external field is increased, some spins flip up, but this process is inhibited by the local exchange field due to the mostly down neighboring spins.  Eventually, some small islands of up spins appear,  and for larger fields, their average size grows. Eventually, when the external field becomes sufficiently large positive, most spins become up. This process of appearance and growth of initially small islands of up spins is well-described by classical nucleation theory  \cite{nucleation, dhar:93}, which we now sketch below. 

Consider an island of up spins immersed in a sea of down spins. There is a free energy cost for creating this island, and the effective free energy for creating a droplet of radius $R$  is given by 
\begin{equation}
\Delta G \approx  A R^{d-1} - B h R^d,
\end{equation}
 where $A$ and $B$ are constants that take care of numerical constants like $\pi$ and temperature-dependent coefficient of surface tension, etc.
The first term takes into account the effect of surface energy that varies as $R^{d-1}$, and the second term describes the decrease in the free energy due to the alignment of the spins to the external (positive) field.  This free energy initially increases for small $R$, and then has a maximum at $R = R^* = ( d-1)A/(hBd)$, and then decreases for larger $R$. Droplets of size $R < R^*$  would generally shrink in size, and disappear, and only if their size becomes greater than $R^*$ would they grow on average.  The process of forming a droplet of size greater than $R^*$ involves an activation barrier $\Delta G^* \approx A_1/ h^{d-1}$, and hence we get the rate of formation of the drops per unit volume $\Gamma$ varies with $h$ as 
\begin{equation}
\Gamma \sim \exp{\left[ - \beta A_1/ h^{d-1} \right]}.
\end{equation}
Once the critical droplet is formed, its radius increases with time, with the radial velocity  of the front $ V_\mathrm{drop}$  proportional to the instantaneous field $h$. Then the order of magnitude time $t_\mathrm{coer}$   at which the net magnetization becomes zero can be estimated by requiring that the density of critical droplet forms by time  $ t= t_\mathrm{coer}/2$  is big enough, so that by the time $t= t_\mathrm{coer}$,  their sizes increase to fill most of the space. Since the dependence of $\Gamma$ on $h(t)$ is very steep, most of such droplets form near the end of the interval. Then, the field at that time is approximately $h_0 \omega t_\mathrm{coer} = h_\mathrm{coer}$. The average volume of a drop  $v$ at a time $t$, with  $t > t_\mathrm{coer}$  grows in time as $  v \sim V_\mathrm{drop} [ t- t_\mathrm{coer}]^d$.  Here $ V_\mathrm{drop}  \sim h$, and $ t_\mathrm{coer} \sim h_\mathrm{coer}/ h_0 \omega$.  Then the equation determining $h_\mathrm{coer}$ is determined by the condition  that  $v \Gamma$ is of order $1$. Thus,
\begin{equation}
h_\mathrm{coer} ^{2d} \exp( -  \beta A_1/ h_\mathrm{coer}^{d-1} ) \sim ( h_0 \omega)^d. 
\end{equation}
For $h_0 \omega$ very small,  the main dependence of the left hand side on $h_\mathrm{coer}$ comes from the argument of the exponential. This implies that
\begin{equation}
h_\mathrm{coer} \sim   |\log ( h_0 \omega)|^{\frac{-1}{(d-1)}}, \quad\text{for}\quad h_0 \omega \to 0.
\end{equation}

While the arguments for limiting behavior are not rigorous, they only depend on the fact that for small $h$, the nucleation time to form a critical droplet has to vary as $ \exp( - \beta A_1/ h^{d-1})$, and hence the conclusion seems unavoidable.  However, this kind of asymptotic dependence is hard to verify in experiments or in Monte Carlo simulations. Sides, Rikvold, and Novotny made an extensive study of the problem in $d =2$ using extensive simulations in the single droplet and multiple droplet growth regimes  \cite{sides1, sides2a, sides2b, sides2c} and concluded that ``we stress that the asymptotic low-frequency behavior would only be seen for extremely low frequencies". They also note that in cases of high activation barriers, the effect of impurities or boundaries of the system become important. This is also true for the Ising model simulations and experiments on films. We do not discuss these finite-size effects in detail here. 

In the literature,  many different forms have been reported. He and Wang studied hysteresis in a few atomic-layer thick films of Fe/Au(001) films for low values of  $h_0$ and $\omega$, and could fit their data to the area of the hysteresis loop varying as $ h_0^\alpha \omega^\beta$, with $\alpha \approx 0.31±0.05.$ and $\beta=0.59±0.07$~\cite{he-wang}. 
\citet{jiang-yang-wang} in their study of hysteresis in thin cobalt films found $\alpha \approx  \beta \approx 2/3$. 
\citet{suen-erskine} also studied thin iron films  and observed a power law scaling of the hysteresis loop area for over five decades in frequency.  Their measured effective exponents are fairly low values $ \alpha \approx 0.254 \pm 0.003,   \beta = 0.063 \pm  .002$, qualitatively consistent with a weak logarithmic dependence~\cite{suen-erskine}.  \citet{sengupta92} used a cell-dynamics simulation and found clear evidence
for the dynamical transition between the asymmetrical loops at low fields for a given frequency, and symmetrical loops at higher amplitudes of the field, and drew the phase boundary for this transition.  Their numerical estimates for the exponents  are $\alpha \approx  0.47 \pm 0.02$ and $\beta \approx 0.40 \pm 0.01$.  The deviations in these studies from the values expected are presumably due to the fact that $h_0$ and $\omega$ are not in the asymptotic regime. 

\section{Hysteresis in  continuous spin models}
\label{s:continuous-spin}

The case of hysteresis in continuous spin models is somewhat outside the overall aim of this volume, which is  about  Ising models.  However, it seems worthwhile to briefly mention here the case of $n$-vector models with $n>1$, and how the hysteresis scaling there differs from the Ising case. 

Historically, the scaling of the area of the hysteresis loop with the amplitude and frequency of the driving fields was first studied by 
\citet{rao-krishnamurthy-pandit}. They studied
the $n$-vector model with nearest-neighbor ferromagnetic couplings, subjected to a time-varying field. Their Hamiltonian was expressed in terms of the dynamics of an $O(N)$-symmetric $(\Phi^4)$ field theory  subjected to a time-varying field, but  is equivalent to the Hamiltonian 
\begin{equation}
H = - J \sum_{nn} \vec{S}_i\cdot \vec{S}_j ~~ - h_0 \sin(\omega t) \sum_i  \hat{e}_1 \cdot \vec{S}_i.
\end{equation}
Here, $\vec{S}_i$ is the $n$ component vector spin at site $i$, and $\hat{e}_1$ is the unit vector along the direction $1$  in the spin-space.  In the large $n$ limit, the non-linear constraint  $|\vec{S}_i|=n$ may be replaced by a, now time-dependent,    Lagrange multiplier $\lambda(t)$, and we have an effective Hamiltonian 
\begin{equation}
H_\mathrm{eff} = -J \sum_{nn} \vec{S}_i\cdot \vec{S}_j - (\lambda(t)/2) \sum_i \vec{S}_i^2 - h_0 \sin(\omega t) \sum_i  \hat{e}_1 \cdot \vec{S}_i. 
\end{equation}
This is a quadratic Hamiltonian, and is easily solved even for the time-varying case. Assume that the spin evolves according to Langevin dynamics.  The equation of motion for $M_i(t) = \langle \vec{S}_i \cdot \hat{e}_1\rangle$ is 
\begin{equation}
\frac{d}{dt} M_i(t) = \nabla^2 M_i(t) -\lambda(t) M_i(t) + \eta_i(t), 
\end{equation}
where $\nabla^2$ is the lattice Laplacian, and $\eta_i(t)$ are  Gaussian white noise variables satisfying 
\begin{equation}
\langle \eta_i(t) \eta_j(t') \rangle = \delta_{ij} \delta(t-t').
\end{equation}
Here the Lagrange multiplier function $\lambda(t)$ is determined by the self-consistency condition that \begin{equation}
\langle \vec{S}_i^2\rangle_t =n, \quad\text{for all}~i, ~\text{ and all}~t.
\end{equation}
These equations can be solved by Fourier transformation.  For the Fourier component $\langle |S_q|^2\rangle$, one gets an equation of the form 
\begin{equation}
\frac{d}{dt} \langle |S_q|^2\rangle  = - [ q^2 + \lambda(t)] \langle |S_q|^2\rangle, |q| \neq 0,
\end{equation}
and for the $q=0$ mode,
\begin{equation}
\frac{d}{dt} \langle S_{q=0}\rangle = \lambda(t) \langle S_{q=0} \rangle + h_0 \sin \omega t. 
\end{equation}
Solving these equations numerically, the authors studied the variation of the shape of the loop with $h_0$ and $\omega$.  They found that the area of the loop varies as $h_0^{\alpha} \omega^{\beta}$, with  $\alpha \approx 2/3$ and $\beta \approx 1/3$ in the small $h_0$ and small $\omega$ limit. It was shown later in~\cite{dhar:92, dhar:93, somoza:93} that in more than two dimensions ($d>2$), for $n \ge 2$, the area of the hysteresis loop
scales as $(h_0 \omega)^{1/2}$ with logarithmic corrections, at low frequency $\omega$ and low
amplitude $h_0$ of the driving field, indicating $\alpha=\beta=1/2$. At high
frequencies, the area varies as $h_0^2/\omega$. For any $h_0$, there is
a dynamical phase transition separating these two frequency
regimes. Above the critical frequency $\omega(h_0)$, the hysteresis
loop does not possess inversion symmetry. The case of the single-component $\Phi^4$ model in a time varying field has been discussed by \citet{mahato}, and more recently by \citet{scopa}.

\section{The  dynamical phase transition}
\label{s:dpt}

It is useful to start by defining precisely what is meant by a phase transition in a non-equilibrium system, where usual quantities like temperature, free energy, etc are not defined, as the system is not in local thermal equilibrium at any time.  Even in the case of Glauber dynamics for the Ising model in a time-dependent field, the system is coupled to a heat bath at a given temperature,  the system is not in equilibrium, and is always relaxing to equilibrium, and does not have a fixed temperature. In general, we can characterize the state in terms of measurable observables like energy density,  or magnetization density at any time, but thermodynamic quantities such as temperature, entropy, or the Gibbs or the Landau free energy do not have any obvious definitions in  the non-equilibrium context.   

For periodically driven systems, if we make stroboscopic observations of the system only at a fixed value of the phase-angle   $ \phi =  \omega t\,( {\rm mod} ~2 \pi)$, then we see a steady state that is a function of the phase-angle  $\phi$. For this steady state, we can define an effective Hamiltonian $H_\mathrm{eff}(\phi)$, and then the familiar language of phase transitions can be used to discuss the phases and phase transitions in a system governed by $H_\mathrm{eff}(\phi)$ \cite{dutta}.  We note that in many cases, $H_\mathrm{eff}(\phi)$ so defined usually has long-range couplings, and discussion of universality classes needs care.

In the present context,  we have a periodic macro-state that is specified by the values of controllable parameters.  If we change the parameters, the macro-state changes.  In the specification of the macro-state, we can ask some yes/no questions. For example, does the hysteresis loop have the inversion symmetry  $ M(-h) =-M(h)$?  In some other dynamical phase transition, we may ask if there is a laser output in an optical pumping setup at a given intensity of the pump laser.   As the control parameters vary continuously, this answer usually remains the same. When it suddenly changes from yes to no (or vice versa), we say that the system has undergone a transition from the `yes' phase to the `no' phase.  In the present case, the yes/no question is ``Is the cycle-averaged magnetization in the periodic steady state zero?".  This definition of a dynamic phase transition is a generalization of the definition of the equilibrium phase transitions. In fact, the static transitions is a special limit of the dynamical transitions, an endpoint of a line of dynamical phase transitions,   where the rate of variation of the control parameters tends to zero. 

\subsection{Dynamical phase transition in the Ising case}

This dynamical phase transition in the context of the Ising model was first studied by \citet{tome}.  They considered the evolution of a set of Ising spins where each spin is coupled to every other with equal strength $J/N$,  evolving  with Markovian  Glauber single-spin flip dynamics.  The Hamiltonian is the time-dependent Hamiltonian 
\begin{equation}
H = - \frac{J}{N} \sum_{i,j} S_i S_j  - ( h_b + h_0 \sin ~\omega t ) \sum_i S_i.
\end{equation}
For this Hamiltonian, the time-dependent  expectation values $\langle S_i\rangle$ satisfy a simple differential equation 
\begin{equation}
\tau \frac{d}{dt} \langle S_i\rangle = -\langle S_i\rangle + \tanh \beta \bigl(J \langle S_i\rangle + h_0 \cos \omega t\bigr).
\end{equation}
This is a non-linear equation, but it is easily solved numerically. 

\begin{figure}
\centering
\includegraphics[width =.6\linewidth]{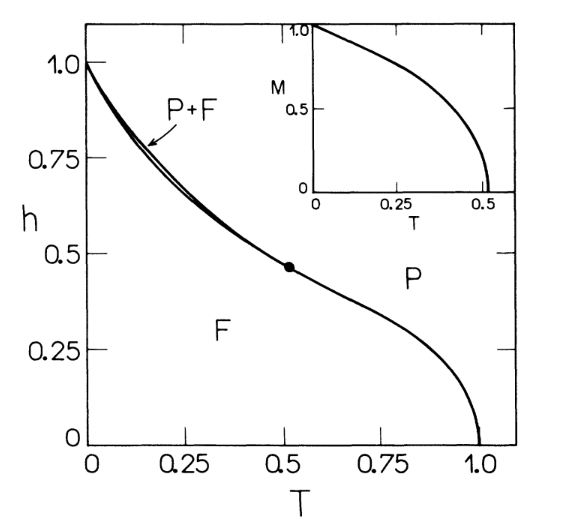}
\caption{Non-equilibrium phase diagram of the Ising model in an oscillating field in the $h-T$ plane, for fixed values of $h_0$ and $\omega$.  
Reprinted figure with permission from~\cite{tome}~[\href{https://doi.org/10.1103/PhysRevA.41.4251}{T. Tom\'e and M.J. Oliveira, Phys. Rev. A {\bf 41}, 4251 (1990)}]. Copyright (1990) by the
American Physical Society.}
\label{fig:tome}
\end{figure}

Tom\'e and Oliveira found that the long-time steady state always settles into a periodic solution. They could classify these into two qualitatively different types: those for which cycle-averaged magnetization $Q$ is zero and non-zero. In this case, the control parameter space is four-dimensional, but for small $h_0$, these are continuously related to the $h_0=0$ thermal equilibrium state. The phases may be identified as the paramagnetic and ferromagnetic phases. In \fref{fig:tome}, we show a two-dimensional cut of the 4 -dimensional non-equilibrium phase diagram in the variables $h_0$ and temperature $T$. Then, the non-equilibrium ferromagnetic/ paramagnetic phases are a continuation of the corresponding equilibrium phases. 

The authors found that in some region of parameter space, both the solutions with $Q$ zero and non-zero are locally stable to small perturbations. In this regime, both phases
are possible and which one is realized depends on the history of preparation. This region is marked by (P+F) in the figure. On the boundary of the transition line between these phases, for small $h_0$, $Q$ is zero at the boundary, but for larger $h_0$, there is a discontinuity in the value of $Q$ as the solution jumps from the symmetric to the asymmetric solution. This is analogous to a tricritical point in equilibrium transitions, though it is best to avoid this terminology borrowed from equilibrium statistical physics here.

\subsection{ Dynamical transition in the continuous spin models}

In the case of $n$-vector models, with $n>1$, there is also a dynamical transition, but it is qualitatively different. For small $\omega$, the magnetization vector stays aligned with the field direction much of the time, and when the field changes sign, $S_z$ changes from near  $+1$ to near $-1$. On increasing $\omega$, the spins spend a higher fraction of the time in between and is mostly in the transverse direction. The magnetization still shows  a long-range order, but the orientation direction is primarily transverse, with a small oscillating component in the longitudinal direction. 

A later more detailed study by \citet{junier} found that for $n=2$, and $d \geq 2$, there are three different phases possible:  paramagnetic inversion symmetric loops, longitudinal or transverse ferromagnetic order, with the cycle-averaged magnetization in the direction of the field or perpendicular to it, respectively, and a canted phase where there is a mean longitudinal and a transverse magnetization. They found continuous phase transitions, from longitudinal to canted to transverse ordering as the frequency increases.    In the $n \geq 3$  case for all $d>2$, there is only one transition from the inversion symmetric paramagnetic to the transverse-ordered phase on increasing the frequency.  This absence of inversion-symmetry breaking phase for the continuous spin case is also consistent with the result of \citet{paessens}, who studied the hysteresis in the spherical model, which is the $n \rightarrow \infty$ limit of the $n$-vector model, and did not find any dynamical phase transition to an inversion symmetry breaking phase.

\section{Hysteresis in disordered ferromagnet and Barkhausen noise}
\label{s:RFIM}

\begin{figure}
    \centering
    \includegraphics[width=0.8\linewidth]{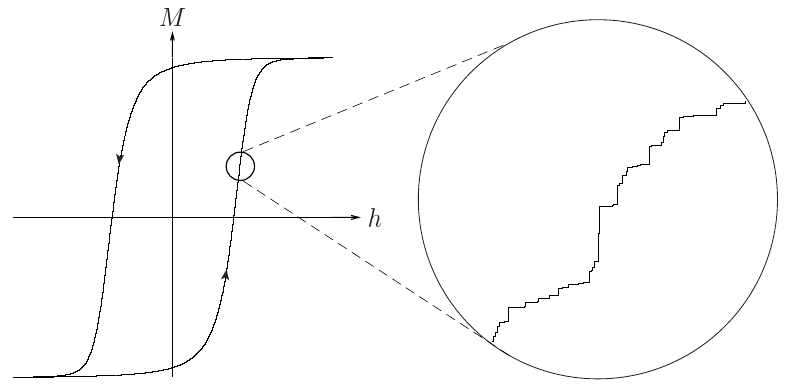}\\
    \vskip7mm
\caption{Hysteresis loop  from a multi-domain sample showing  magnetization jumps. Figure reproduced from the SS’s PhD thesis~\cite{sabhapandit:phd}.} \label{fig:barkhausen1} 
\end{figure}

So far, we have discussed hysteresis in magnets that are  macroscopically spatially homogenous. If we observe the hysteresis loop in a typical piece of iron, it consists of many randomly oriented domains. 
Then, the change in magnetization occurs in many discrete jumps  of a wide range of sizes, corresponding to individual domains changing their orientation. This can be picked up in a microphone and leads to a crackling noise that is called the Barkhausen noise 
~\cite{barkhausen:19, feynman:77, sipahi:94, spasojevic:96} (Fig. \ref{fig:barkhausen1}).  The size of these jumps can vary over three to four orders of magnitude. This is an example of a kind of scale-free (hence `critical') response of a non-equilibrium  driven system that occurs without requiring a fine-tuning of parameters, which  is known as  self-organized criticality \cite{spasojevic:96}.

\begin{figure}
\centering 
    \includegraphics[height=0.34\linewidth]{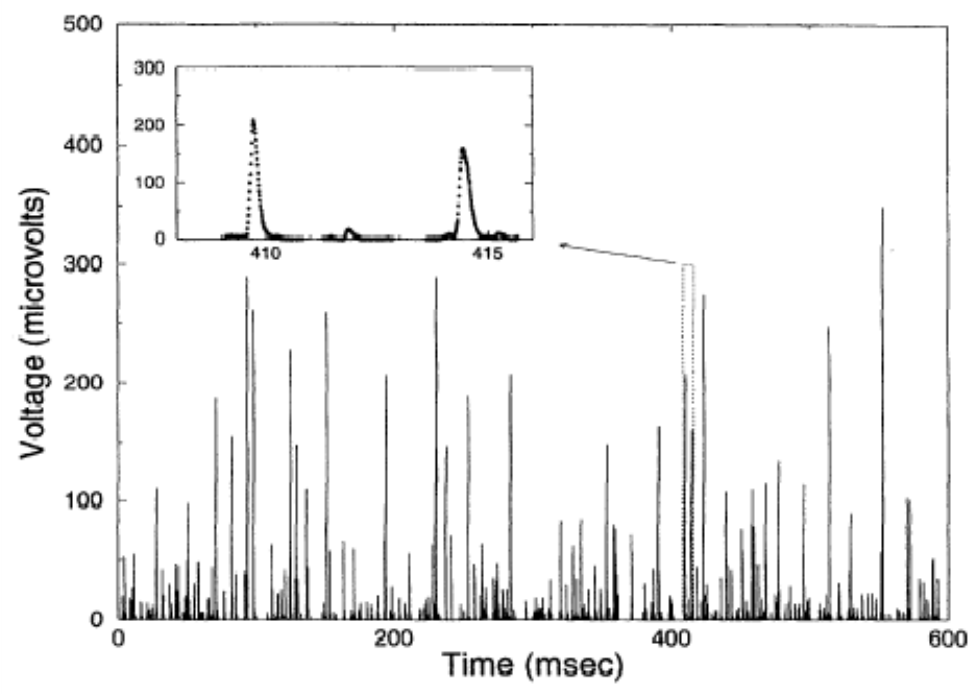}~~\includegraphics[height=0.34\linewidth]{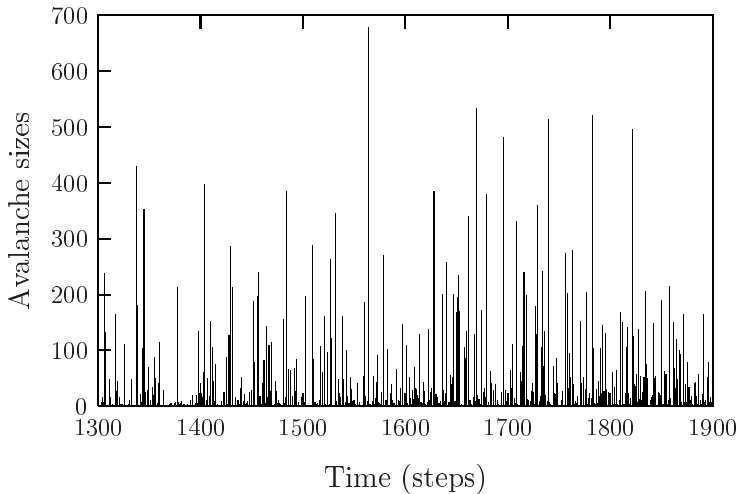}
    \caption{Left:~Experimental Barkhausen signal--- voltage pulse from a pickup coil around a ferromagnet under a slowly varying field. 
Reprinted figure with permission from~\cite{urbach:95}~[\href{https://doi.org/10.1103/PhysRevLett.75.276}{J. S. Urbach,  R. C. Madison, and J. T. Markert,  Phys. Rev. Lett. {\bf 75}, 276 (1995)}]. Copyright (1995) by the
American Physical Society. 
Right:~Time series of avalanches (number of spin flips per field step) in the 
random field Ising model on a $200\times 200$ square lattice. Consecutive 
avalanches define successive time steps. 
The right figure is reproduced from the SS’s PhD thesis~\cite{sabhapandit:phd}.}
\label{fig:barkhausen2}
\end{figure}

Sethna et al. proposed a simple model that shows  this phenomenology: the random field Ising model (RFIM) at zero temperature.
The behavior of the random field Ising model in thermal equilibrium is not fully understood. 
\citet{imry:75} argued that
arbitrarily weak disorder destroys the long-ranged ferromagnetic order in
dimensions $d\le2$. Later, \citet{imbrie:84} showed that if the disorder is small, the
model in three dimensions ($d=3$) exhibits long-range order at zero
temperature. \citet{aizenmann:89} rigorously proved the uniqueness of the
Gibbs state in two dimensions ($d=2$),  indicating the absence of any phase transition, in
agreement with the Imry--Ma argument.

As far as an exact calculation of thermodynamic quantities is concerned, there are only a few results. 
There are no known exact results for the average
free energy or magnetization for a continuous distribution of random
field, even at zero temperature and in zero applied field. A comprehensive summary covering examples of interesting realizations in nature, equilibrium properties for both short-range and long-range correlated random fields, interface properties of domain walls, the nature of the phase transition, scaling laws and critical exponents, as well as the zero-temperature dynamics of domain walls near the depinning transition in the presence of an external field, along with references to several earlier reviews, can be found in the review by~\citet{nattermann1998theory}.

For the hysteretic response of the RFIM, the problem can be solved exactly in one dimension \cite{kurbah:2011}. 
For the two-dimensional case, there are only numerical simulations.  See for example, \cite{salvat-pujol:2009}.  Even though the equilibrium properties of the RFIM model are not easy to determine exactly, it turns out that the non-equilibrium  hysteretic response can be determined exactly in some cases.

In the case of the Ising model without disorder, we have seen that the hysteresis is governed by the rate of formation and growth of droplets of the oppositely magnetized phase.  The activation barriers are strongly modified by the presence of impurities. We now discuss the case of hysteresis in a random field Ising model, where the effects of disorder are predominant, and a large number of droplets are present even in ground state, which qualitatively change the behavior of the hysteresis loops.

 In this model, an Ising spin $s_i=\pm 1$ is placed on each site of a lattice and nearest-neighboring spins interact ferromagnetically. Each spin is also coupled to an on-site 
quenched random field $h_i$, drawn independently from a common distribution $\phi(h_i)$ at each  site of a lattice. In addition, the entire system is subjected to a spatially uniform time-dependent magnetic field $h(t)$, which can be varied externally. For a given configuration of the spins $\{s_i\}$, a given realization of the quenched random fields $\{h_i\}$, and a given value of the external field $h$, the energy of the system of given by
\begin{equation}
    E= -J\sum_{\langle i,j\rangle} s_is_j -\sum_i h_i s_i -h(t) \sum_i s_i\, .
    \label{eq:rfimE}
\end{equation}
At a fixed value of the external field $h$, the system relaxes through zero-temperature Glauber single-spin-flip dynamics, whereby a single spin flips at a certain rate $\Gamma$ only if the spin-flip lowers the energy of the system. Because of the energy-lowering dynamics, at any $h$, the system eventually always relaxes to a stable spin configuration, in which the spins are aligned to the local field $\ell_i(h) = J \sum_{j \in \partial i} s_j + h_i + h$, i.e., 
\begin{equation}
    s_i = \sgn \left( \ell_i(h)\right) = \sgn\left(J \sum_{j \in \partial i} s_j + h_i + h\right),
\end{equation}
where the sum runs over the nearest neighbors $\partial i$ of site $i$. 
If we start from a stable spin configuration at a given external field $h$, 
and then change the field to a different value $h'$, the local fields change 
from $\{\ell_i(h)\}$ to $\{\ell_i(h')\}$. The configuration remains stable if  
$
\sgn\!\big(\ell_i(h')\big) = \sgn\!\big(\ell_i(h)\big) $
for all $i$.
On the other hand, if for some sites the local field change sign,  
$\sgn\!\big(\ell_k(h')\big) = -\,\sgn\!\big(\ell_k(h))$, then the corresponding spins become unstable. Suppose $h'$ is 
such that only a single site $k$ meets this condition. The spin at that site 
flips, $s_k \to -s_k$. This flip may in turn render neighboring spins unstable 
by changing the sign of their local fields, triggering further flips. The resulting 
cascade continues until all spins once again become locally stable. Thus, the flip of a single spin may trigger an avalanche of spin flips. Because of this, when the external magnetic field $h$ is varied adiabatically 
(i.e., at a rate much slower than $\Gamma$, so that $h$ remains effectively 
constant during an avalanche), the magnetization changes occur in discrete 
jumps corresponding to avalanches of various sizes (see \fref{fig:barkhausen1}). This provides a simple theoretical analog
of the so-called Barkhausen noise, which is heard in a speaker attached to a pickup coil wound around a ferromagnetic sample~\cite{barkhausen:19, feynman:77, sipahi:94, spasojevic:96} (see ~\fref{fig:barkhausen2}). 

Note that the adiabatic limit may be equivalently realized by assuming that unstable spins flip instantaneously, i.e., $\Gamma \to \infty$. Therefore, by adiabatic variation of the external field $h(t)$, we mean that the system relaxes 
instantaneously to a stable configuration at each value of the field.
The zero temperature random field Ising model with adiabatic variation of the external field possesses two important properties, namely the \emph{return point memory effect}~\cite{sethna:93} and the \emph{abelian nature of spin flips}~\cite{dhar:97}, which we briefly discuss below.

\emph{Return point memory}:~(\citet{sethna:93}; also see also~\cite{sabhapandit:phd}) Suppose we start with $h=-\infty$ with all spins down at $t=0$, and increase the 
field adiabatically such that $h(t) \leq h(\mathcal{T})$ for all $t<\mathcal{T}$. 
Then the spin configuration at $t=\mathcal{T}$ depends only on the maximum field 
$h(\mathcal{T})$ and not on the detailed history of $h(t)$. If 
this maximum was already attained at some earlier time $t'$, the configuration 
(and hence the magnetization) at $t=\mathcal{T}$ is identical to that at $t'$.

\emph{Abelian property}~\cite{dhar:97}: Owing to the above property, the external field may equivalently be 
raised in a single step from $-\infty$ to $h(T)$. At this value, several 
spins can acquire positive local fields. If multiple such unstable sites 
exist, flipping one of them necessarily increases the local field at the 
others, since all couplings are ferromagnetic. Consequently, all unstable 
spins must eventually flip, leading to a unique stable configuration that 
is independent of the relaxation order. By spin-reversal symmetry, the 
abelian property applies equally when the field changes from $h'$ to $h''$, 
regardless of whether $h''>h'$ or $h''<h'$, provided both configurations 
are stable.

\subsection{Mean-field theory}

\begin{figure}
    \centering
    \includegraphics[width=0.33\linewidth]{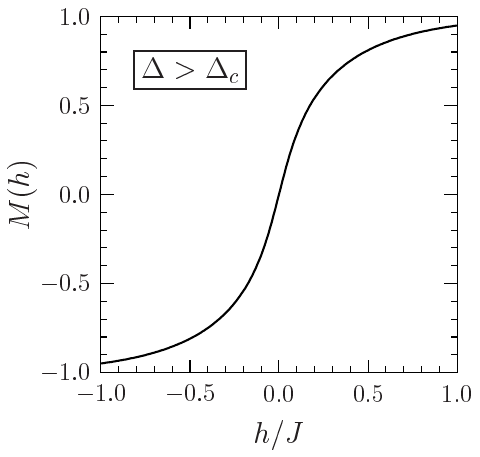}~~\includegraphics[width=0.33\linewidth]{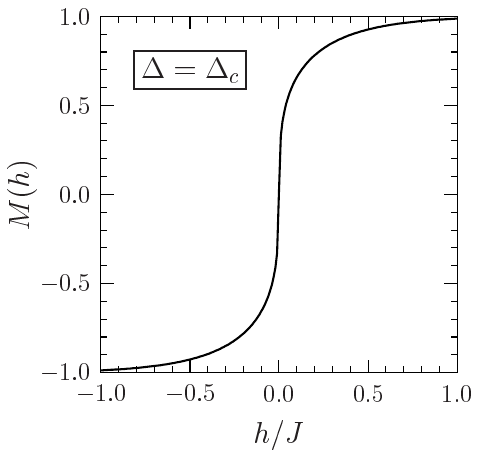}~~\includegraphics[width=0.33\linewidth]{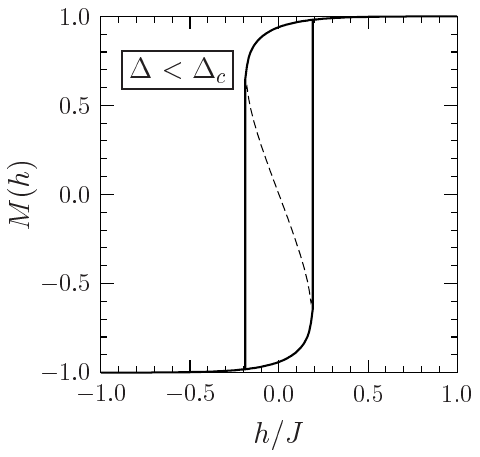}
    \caption{\label{mf-mag} 
Magnetization curves of the random field Ising model with infinite-range 
interactions for different values of the disorder strength: 
(left) $\Delta=J >  \Delta_c$, (middle) $\Delta=\Delta_c=\sqrt{2/\pi}\,J$, and 
(right) $\Delta=0.5J <\Delta_c$, for the Gaussian random field distribution $\phi(h_i) = e^{-h_i^2/(2\Delta^2)}/\sqrt{2\pi \Delta^2}$. The dashed line in the right panel indicates the third solution 
of the self-consistent equation~(\ref{eq:self-consistent}). Reproduced from SS’s PhD thesis~\cite{sabhapandit:phd}.}

    \label{fig:MFhys}
\end{figure}

\citet{sethna:93} studied the hysteresis of the random field Ising model within mean-field theory, 
where each spin is coupled to all other spins with a coupling constant $J/N$. In the $N\to\infty$ limit, 
 the system averaged magnetization $\bigl(\sum_{i=1}^N s_i\bigr)/N$ converges to the statistically averaged magnetization $M \equiv \langle s_i \rangle$,
 yielding $E= - \sum_i (JM + h+ h_i) s_i$ and $s_i = \sgn(JM + h+ h_i)$. Therefore, the magnetization is self-consistently determined via
\begin{equation}
    M(h) = \langle \sgn(JM + h+ h_i)\rangle =2 \int_{-JM(h)-h}^{\infty} \phi(h_i) \, dh_i -1,
    \label{eq:self-consistent}
\end{equation}
where we recall that $\phi(h_i)$ is the distribution of the random fields. If $\phi(h_i)$ is symmetric, then $\int_{0}^{\infty} \phi(h_i) \, dh_i = 1/2$, and clearly, $M(0)=0$ is a trivial solution at $h=0$.
The existence of a hysteresis loop depends on whether~\eref{eq:self-consistent} admits other nontrivial solutions at $h=0$ or not~(see \fref{fig:MFhys}). 

The trivial solution for small $h$ is given by $M(h) = M'(0) h +O(h^2)$ at the leading order in $h$. Using this in~\eref{eq:self-consistent} and evaluating the integral to the linear order $h$, we get $M'(0)= 2\phi(0)/[1-2J\phi(0)]$. If $M'(0) >1$, i.e., $\phi(0) < 1/(2J)$, then \eref{eq:self-consistent} admits a unique solution and there is no hysteresis.  On the other hand, if $\phi(0) > 1/(2J)$, the slope $M'(0)$ at $h=0$ is negative (see the dashed line in \fref{fig:MFhys} for $\Delta< \Delta_c)$, leading to at least two additional nontrivial solutions, and hence hysteresis. 

For a Gaussian distribution $\phi(h_i) = e^{-h_i^2/(2\Delta^2)}/\sqrt{2\pi \Delta^2}$ of random fields, it corresponds to a critical value of the disorder strength $\Delta_c = J\sqrt{2/\pi}$ such that the hysteresis loop exists only for $\Delta < \Delta_c$ (see \fref{fig:MFhys}). In fact, for the Gaussian random fields, the self-consistent equation~(\ref{eq:self-consistent}) is explicitly given by
\begin{equation}
    M(h) = \text{erf}\left(\frac{J M (h) + h}{\sqrt{2 \Delta^2}  }\right).
\end{equation}

A major drawback  of infinite-ranged interaction approximation is the absence of hysteresis above a 
critical disorder. Furthermore, below the critical disorder, when hysteresis 
does occur, it is invariably accompanied by an infinite avalanche at some 
value of the external field. These shortcomings of infinite-range interactions 
in mean-field theory can be overcome by considering the model on a Bethe lattice, 
which incorporates correlations through nearest-neighbor spin interactions. 
At the same time, the Bethe lattice retains certain mean-field-like features: The closure approximation becomes exact on the Bethe lattice, as there are no short loops.

\subsection{Hysteresis in the random field Ising model on a Bethe lattice}
\label{s:rfim-hys-Bethe}

We now discuss the  exact solution of the random field Ising model on the Bethe lattice~\cite{dhar:97}.  We give a brief outline of the arguments below and refer to~\cite{dhar:97} for details. 

\begin{figure}
    \centering
    \includegraphics[width=0.8\linewidth]{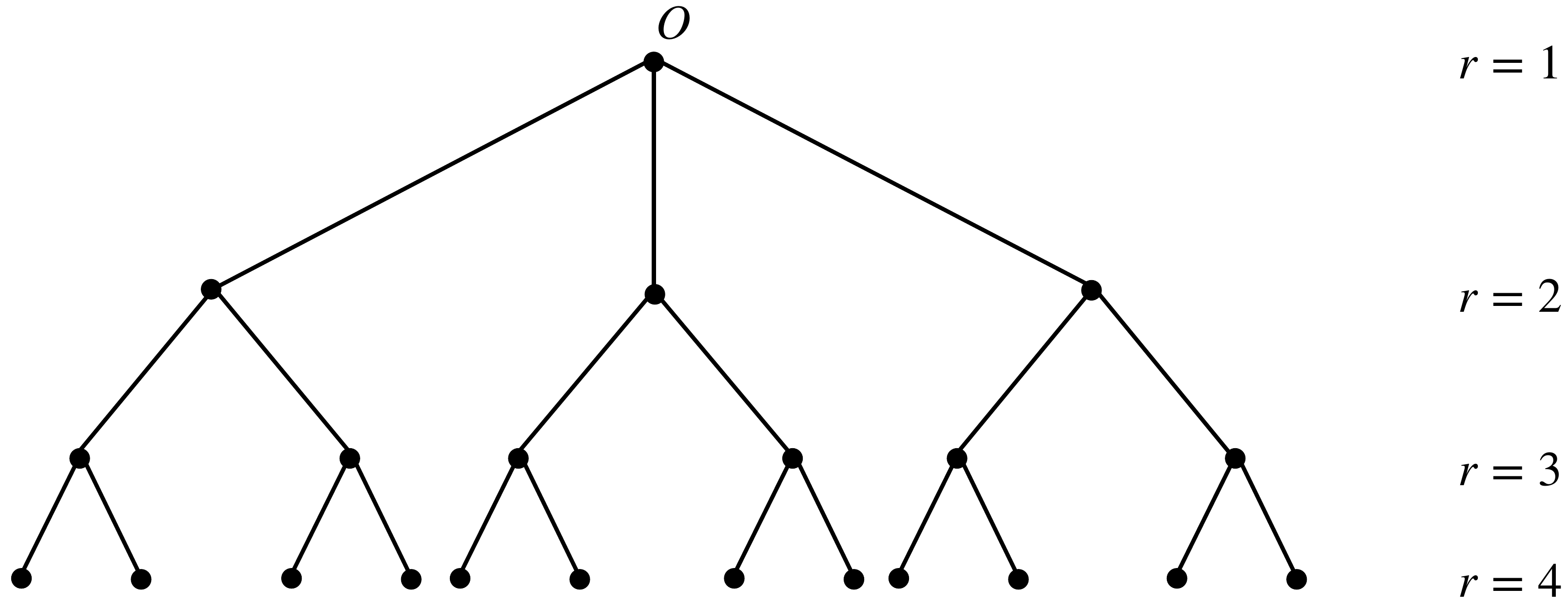}
    
    \caption{A Cayley tree of coordination number $z=3$ and  $n=4$ generations. The Bethe lattice is the deep interior part of a Cayley tree ($r\ll \infty$) when the number of generations  $n\to \infty$.}
    \label{fig:cayley}
\end{figure}

The Bethe lattice can be constructed by considering the deep interior part of a Cayley tree when the number of generations tends to $\infty$ (see \fref{fig:cayley}). 
Consider a uniform Cayley tree of $n$ generations, in which every 
non-boundary site has coordination number $z$, whereas the boundary 
sites (leaf nodes) have coordination number $1$ (see \fref{fig:cayley}). The first 
generation contains a single root vertex, and the $r$-th generation 
consists of $z(z-1)^{r-2}$ vertices for $r \geq 2$. Each vertex carries an Ising spin and a quenched random field, and the 
energy of the system is given by~\eref{eq:rfimE}.
Since all sites of the Bethe lattice far from the boundary  are equivalent, one may choose any one of them to find the average magnetization. We  choose the root $O$ of the tree (see \fref{fig:cayley}), and take the limit $n \to \infty$. 

The hysteresis loop consists of two magnetization curves: a lower curve 
$M_l(h)$ obtained when the external field $h$ is increased from $-\infty$ 
to $\infty$, and an upper curve $M_u(h)$ obtained when $h$ is decreased 
from $\infty$ to $-\infty$. The two curves are related by the symmetry 
$M_u(h) = -M_l(-h)$; hence it suffices to determine only one of them. 
We focus on the lower curve $M_l(h)$. Starting with $h=-\infty$ and all 
spins down, the return point memory property implies that $M_l(h)$ at a given field $h$ can be obtained by raising the field directly from 
$-\infty$ to $h$. Although several spins may become unstable at this 
new field, the abelian property allows them to be flipped in any order. 

For convenience, we relax the spins generation by generation: beginning 
with the leaf nodes (the $n$-th generation), then proceeding to the 
$(n-1)$-th generation, and so on. In general, once all spins at the 
$r$-th generation are relaxed, we proceed to those at the $(r-1)$-th 
generation. While flipping a spin at generation $r$ may destabilize a 
previously stable spin at $(r+1)$, the tree structure ensures that such 
flips cannot influence spins at $(r-1)$. Consequently, when computing the magnetization at the root site $O$, once 
all spins at the $r$-th generation are fully relaxed, there is no need to 
return to the $(r+1)$-th or higher generations.
This hierarchical relaxation greatly simplifies the dynamics: starting 
from the boundary, one proceeds inwards generation by generation until 
reaching the root. In this relaxation scheme, spin flips within the same generation $r$ are mutually independent. In fact, because of this simplification, we are able to calculate the non-equilibrium hysteresis response on the Bethe lattice exactly, but the  equilibrium behavior of RFIM  is much more complicated, and has not been determined  so far.

Let $P_r(h)$ denote the conditional probability that a spin at the 
$r$-th generation flips up, given that all its direct descendants at the 
$(r+1)$-th generation are fully relaxed and the parent spin at the 
$(r-1)$-th generation is kept down. The probability that exactly $m$ of the 
$(z-1)$ descendant spins are up is 
$\bigl[P_{r+1}(h)\bigr]^m \, \bigl[1-P_{r+1}(h)\bigr]^{z-1-m}$, and these 
$m$ spins can be chosen in $\binom{z-1}{m}$ distinct ways. 
Given that there are exactly $m$ up and $(z-m)$ down neighbors 
$(z-1-m)$ down descendants and one down parent), the probability that the 
$r$-th generation spin flips up is equal to the probability that its local 
field $[(2m-z)J + h +h_i]$ is positive. This condition is equivalent to requiring that the random field at the site 
satisfies $h_i > (z-2m)J - h$, which occurs with the probability 
\begin{equation}
p_{m}(h)=\int_{(z-2m)J-h}^{\infty} \phi(h_{i})\, dh_{i}.
\label{p_m}
\end{equation}
Combining these ingredients, we obtain the recursion relation for $P_r(h)$ 
in terms of $P_{r+1}(h)$ as
\begin{equation}
    P_r(h) = \sum_{m=0}^{z-1} \binom{z-1}{m}\, 
    \bigl[P_{r+1}(h)\bigr]^m \, \bigl[1-P_{r+1}(h)\bigr]^{z-1-m}\, p_m(h)\, .
    \label{Pr}
\end{equation}

For a given field $h$, the quantity $p_m(h)$ is obtained from \eref{p_m}. Substituting into \eref{Pr} and using the boundary 
condition $P_n(h)=\int_{J-h}^\infty \phi(h_i)\, dh_i$, one can evaluate $P_r(h)$ for all $r<n$. 
Deep inside the tree ($r \ll n$), $P_r(h)$ converges to a fixed-point 
value $P^\star(h)$, determined self-consistently by  
\begin{equation}
P^\star(h)= \sum_{m=0}^{z-1} \binom{z-1}{m} 
\bigl[P^\star(h)\bigr]^m \bigl[1-P^\star(h)\bigr]^{z-1-m}\, p_m(h).
\label{P*}
\end{equation}
\Eref{P*} is a polynomial of degree $z-1$ that determines $P^\star(h)$ in terms of the set $\{p_m(h)\}_{m=0}^{z-1}$.  
In turn, the probability that the root spin $O$ is up can be expressed as  
\begin{equation}
\prob(s_O=+1 \mid h) = \sum_{m=0}^{z} \binom{z}{m} 
\bigl[P^\star(h)\bigr]^m \bigl[1-P^\star(h)\bigr]^{\,z-m}\, p_m(h).
\label{P(O)}
\end{equation}
The lower magnetization curve on the Bethe lattice is
\begin{equation}
    M_l(h) = 2\,\prob(s_O=+1 \mid h) - 1 \, ,
    \label{eq:M}
\end{equation}
while the upper branch follows from the symmetry relation 
$M_u(h) = -M_l(-h)$.

\begin{figure}[t]
    \centering
    \includegraphics[width=0.48\linewidth]{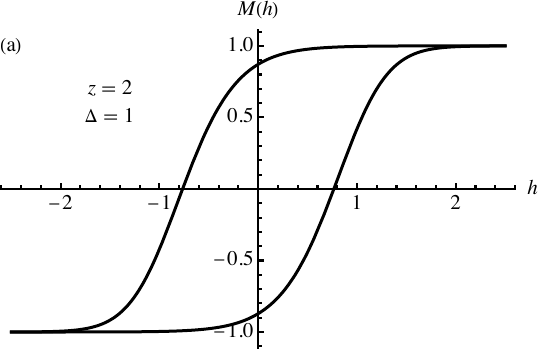}~~~~\includegraphics[width=0.48\linewidth]{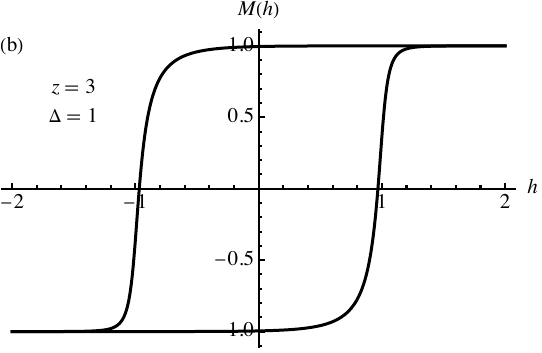}
    \caption{Hysteresis loops for (a) $z=2$ and (b) $z=3$, for Gaussian distribution of random fields with the variance $\Delta^2=1$ and $J=1$. }
    \label{fig:hys2-3}
\end{figure}

\begin{figure}[t]
    \centering
\includegraphics[width=.3\hsize]{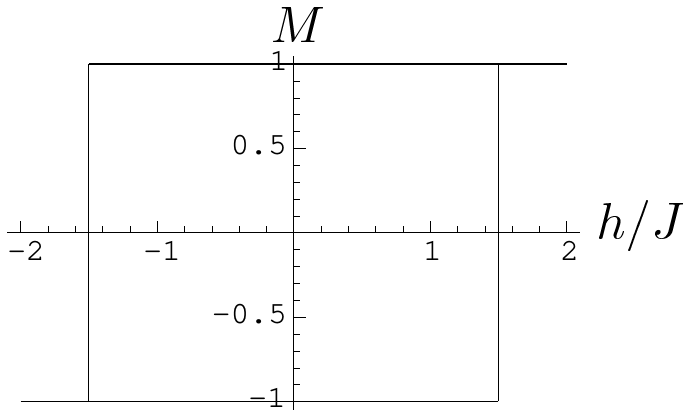}~~~
{\includegraphics[width=.3\hsize]{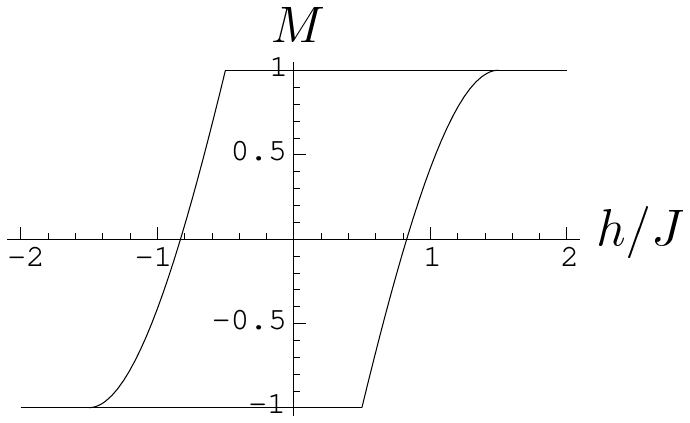}}~~~
\includegraphics[width=.3\hsize]{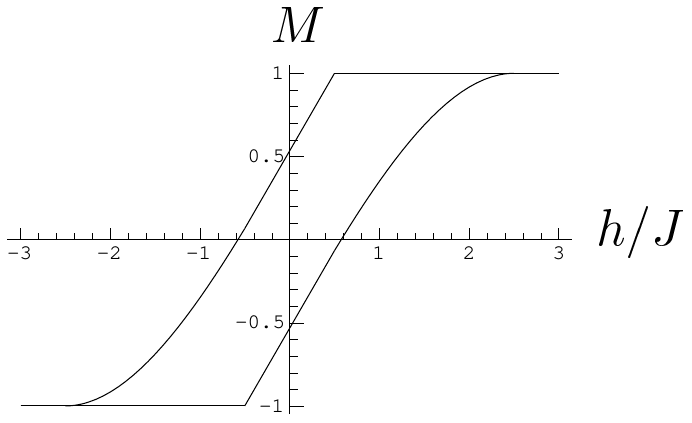}
\caption{Hysteresis loops for a linear chain ($z=2$)  for the rectangular
distribution of quenched random fields in $h_i\in [-\Delta, \Delta]$ with different widths: 
$\Delta/J=0.5$ (left), $\Delta/J=1.5$ (middle), and $\Delta/J=2.5$ (right). (Taken from \cite{sabhapandit:phd})}
\label{fig:m2}
\end{figure}

For $z=2$, the Bethe lattice reduces to a linear chain, where \eref{P*} is linear and yields
$P^\star(h) = {p_0(h)}/[{1-(p_1(h)-p_0(h))}] $.
For $z=3$, \eref{P*} becomes a quadratic. In this case,  between the two solutions, the physical one is the root satisfying $P^\star(h)\in(0,1)$. In both cases ($z=2,3$), for continuous distribution of random fields with unbounded support,  $P^\star(h)$ varies smoothly from $0$ to $1$, resulting in continuous hysteresis loops without any jump discontinuity (see \fref{fig:hys2-3}). It is, however, possible to have jump discontinuity in the magnetization for random field distributions with bounded supports, even for $z=2, 3$. For example, for a linear chain ($z=2$), for a uniform distribution of random fields within the bounded supports $h_i \in [-\Delta, \Delta]$, the magnetization jumps from $-1$ to $+1$ (and vice versa) in a single infinite avalanche for $\Delta < J$, thereby, generating a rectangular hysteresis loop (see \fref{fig:m2}).  

Therefore, for both $z=2$ and $3$, \eref{P*} admits a unique physical solution $P^\star(h)\in (0,1)$. Moreover, for a continuous distribution of random fields with unbounded support, the physical solution $P^*$ varies continuously and monotonically with $h$, leading to a smooth magnetization curve as a function of $h$. 

\begin{figure}[t]
    \centering
    \includegraphics[width=0.33\linewidth]{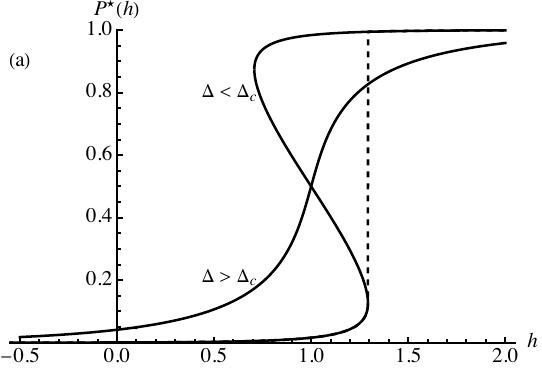}~~\includegraphics[width=0.33\linewidth]{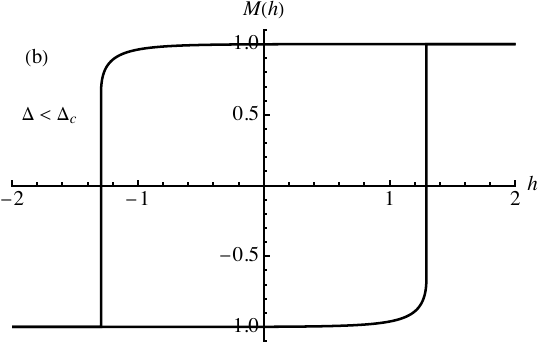}~~\includegraphics[width=0.33\linewidth]{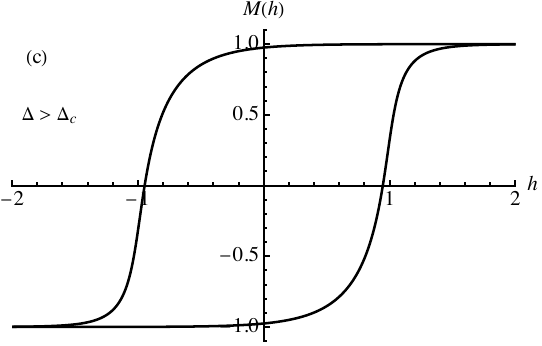}
    \caption{Hysteresis for Bethe lattice with coordination number $z=4$: (a) The real roots of the cubic in the physical interval \eref{P*} $P^\star(h) \in (0,1)$   are plotted for Gaussian distribution $\phi(h_i) = e^{-h_i^2/(2\Delta^2)}/\sqrt{2\pi \Delta^2}$ of the random fields for both $\Delta < \Delta_c$ and $\Delta >\Delta_c$.  Corresponding hysteresis loops for (b) $\Delta< \Delta_c$ and (c) $\Delta>\Delta_c$.   }
    \label{fig:hys4}
\end{figure}

However, this situation changes for $z\ge 4$. For example, for $z=4$, \eref{P*} is a cubic equation that may admit either one or three real roots within the physical interval $P^\star \in (0,1)$ [see \fref{fig:hys4}~(a)]. It turns out that for disorder strength $\Delta$ larger than a critical value $\Delta_c$, the physical root $P^\star(h)$ is a continuous,  monotonically varying function of $h$. Consequently, the hysteresis loop does not exhibit any jump discontinuity for $\Delta >\Delta_c$ [see \fref{fig:hys4}~(c)]. In contrast, for $\Delta<\Delta_c$, starting with $h=-\infty$ and all spins down,  \eref{P*} has a single real root in the interval $(0,1)$ for large negative values of the field $h$. As the field is increases, two additional real roots emerge in the interval $(0,1)$. However, these two new roots lie above the existing root, and $P^\star(h)$ still continues to follow the original branch on physical grounds. Upon further increasing $h$, at a characteristic field  $h_\mathrm{coer}$, the lower and the middle roots merge to become imaginary and disappear from the real plane. At that point,  $P^\star$ jumps to the upper value, giving rise to a jump discontinuity in the corresponding magnetization curve
[see \fref{fig:hys4}~(b)].
This behavior generalizes to higher coordination numbers $z>4$, where the mechanism remains the same: two real roots of Eq.~(\ref{P*}) coalesce to become complex, leading to jump discontinuities in the magnetization.

\begin{figure}
    \centering
    \includegraphics[width=0.8\linewidth]{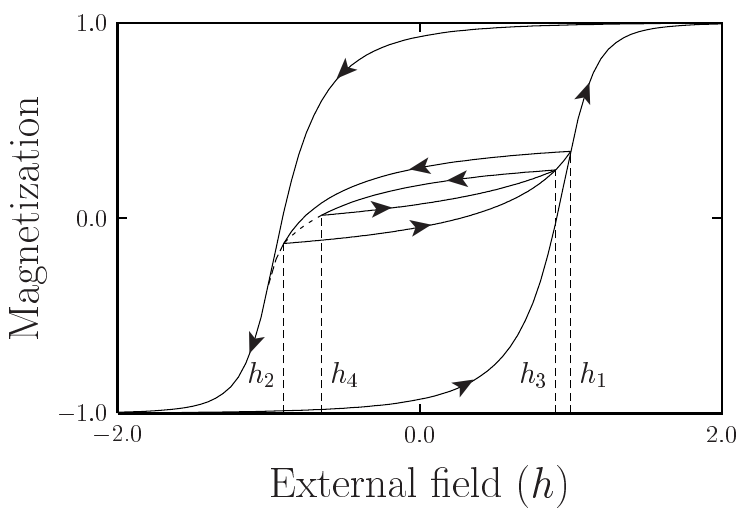}
    \caption{Minor hysteresis loops for Bethe lattice. Reproduced from the author’s PhD thesis~\cite{sabhapandit:phd}.}
    \label{fig:minorloop}
\end{figure}

The hysteresis loop obtained by increasing $h$ from $-\infty$ to $\infty$, and then reversing it back to $-\infty$, is referred to as the ``major'' hysteresis loop. Suppose the system is on the lower hysteresis branch at some external field $h_1$, i.e., starting from $h=-\infty$ with all spins down and increasing the field up to $h_1$. If the field is then decreased from $h_1$ to $h_2$ and subsequently increased back to $h_1$, one obtains the first minor loop (see \fref{fig:minorloop}). Similarly, starting from this first minor loop at some field $h_3$, decreasing the field to $h_4$, and then increasing it back to $h_3$ generates the second minor loop, and so on. In general, the $n$-th minor loop ($n>1$) is obtained from the lower half of the $(n-1)$-th minor loop by decreasing the field from $h_{2n-1}$ to $h_{2n}$ and then increasing it to $h_{2n+1}<h_{2n-1}$. This construction involves the set of turning points $\{h_n\} \equiv \{h_n,h_{n-1},\ldots,h_1\}$, which encodes the full history of field reversals up to $h_n$.

The framework developed for calculating the major hysteresis loop on the Bethe lattice can be extended to determine the minor loops as well. The essential methodology remains similar, but the analysis becomes 
considerably more involved since one has to carefully track the entire history of the field reversals and turning points. A detailed discussion of these calculations lies beyond the scope of the present review, and we instead refer the interested reader to Refs.~\cite{sabhapandit:phd,shukla:00,shukla:01}.

\section{Distribution of avalanche sizes in the random field Ising model on Bethe lattice} 
\label{s:avalanche}

While the disorder-averaged magnetization curve appears smooth 
(apart from the macroscopic jump discontinuity for $\Delta < \Delta_c$), 
for any fixed realization of the random fields $\{h_i\}$ the magnetization 
evolves through discrete jumps of various avalanche sizes, as illustrated 
in \fref{fig:barkhausen1} and discussed in \sref{s:RFIM}. 
The distribution of avalanche sizes on the Bethe lattice was obtained in \cite{sabhapandit:00}. We present only a brief sketch of the main results, and omit the technical details.

Let us start with $h=-\infty$, where all spins are down. 
Next, increase the field adiabatically to a value $h$ and relax all the spins completely. 
Now, increase the field by a small amount $dh$ such that exactly one spin in the system becomes unstable. 
Since all sites of the Bethe lattice are equivalent, we may, without loss of generality, 
assume this unstable spin to be at the root $O$ in~\fref{fig:cayley}. 
The flip of this unstable spin at $O$ may trigger some of its descendant spins at generation $r=1$ to flip, 
which, in turn, may cause their own descendants to flip, and so on, resulting in an avalanche. 
We are interested in determining the probability $G_s(h)\, dh$ of observing an avalanche of size $s$ initiated at $O$, when the field is increased from $h$ to $h+dh$.

\begin{figure}
    \centering
    \includegraphics[width=\linewidth]{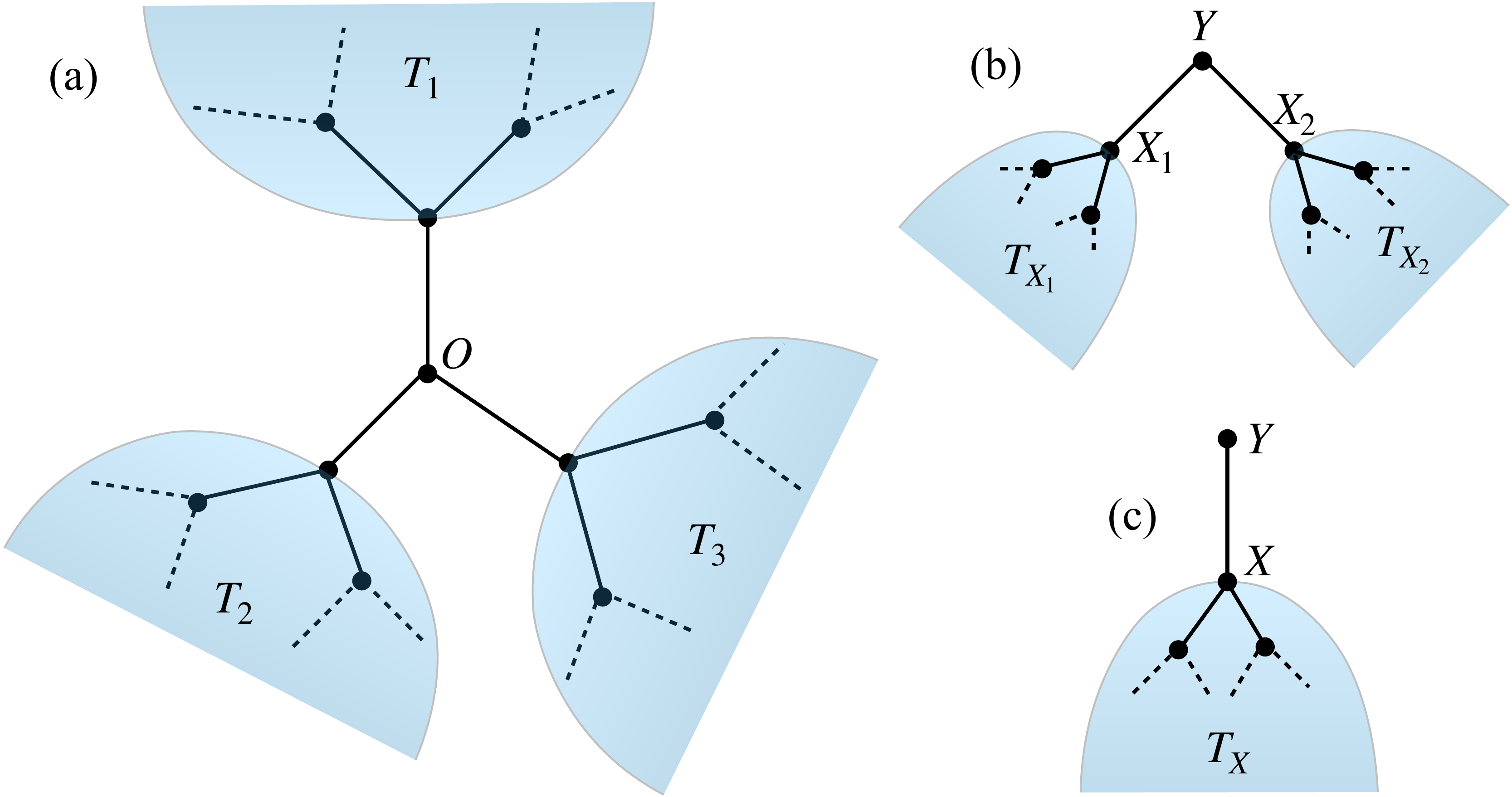}
    \caption{
    (a) A Bethe lattice (deep interior of a Cayley tree shown in \fref{fig:cayley}) with coordination number $z=3$. An avalanche initiated at the vertex $O$ propagates along the $z$ descendants subtrees $\{T_1, T_2, \dotsc, T_z\}$ independently.  (b) Any vertex $Y$ within a specific subtree $T_i$ has its own $z-1$ descendant subtrees $\{T_{X_1},T_{X_2}, \dotsc, T_{X_{z-1}} \}$ rooted at 
    $\{X_1,X_2, \dotsc, X_{z-1}\}$ respectively. (c)
    A subtree $T_X$ rooted at  $X$, consisting of $X$ and all its descendants. The parent of $X$ is denoted by $Y$. }
    \label{fig:sub-tree}
\end{figure}

To obtain the stable spin configuration at field $h$ before the spin at $O$ flips when the field is increased by $dh$, 
we employ the same relaxation scheme described in~\sref{s:rfim-hys-Bethe}: starting from the boundary (leaf nodes) and 
proceeding inward, generation by generation, until reaching generation $r=2$. 
For a completely stable configuration, once a spin at generation $r$ flips, one would, in principle, need to reexamine all of its 
descendant spins at generations $(r+1)$ and beyond for possible flips. 
However, recall from~\sref{s:rfim-hys-Bethe} that this additional reexamination step was unnecessary when determining the 
magnetization. 
Similarly, for avalanche statistics, since an avalanche can propagate outward from the root $O$ only through vertices that are initially in the down state, it suffices to work with this partially relaxed configuration at field $h$. 
In other words, the probability that an avalanche initiated at $O$ has size $s$ is the same in this partially relaxed state as in the fully relaxed state.

The avalanche size distribution $G_s(h)\, dh$ can be expressed in terms of the probabilities of avalanches propagating along the descendant subtrees. 
Let $Q_n(h)$ denote the probability that an avalanche propagating within a subtree flips exactly $n$ additional spins in that subtree before stopping. 
Since, from any vertex $Y$, an avalanche propagates independently along all its descendant subtrees (see \fref{fig:sub-tree}), the corresponding generating function 
\[
   Q(x,h) = \sum_{n=0}^\infty Q_n(h)\, x^n
\]
satisfies the self-consistent equation
\begin{equation}
   Q(x,h) = Q_0(h) + x \sum_{m=0}^{z-1} { {z-1} \choose m }
   [P^\star(h)]^m \,[Q(x,h)]^{\,z-1-m} \,[p_{m+1}(h)-p_m(h)] ,
   \label{Q(x)} 
\end{equation}
where $p_m(h)$ is defined in~\eref{p_m}, and $P^\star(h)$ is the probability that a spin $s_X$ at a vertex $X$ is up in the relaxed configuration at field $h$, given that its parent spin $s_Y$ at $Y$ is kept down (see \fref{fig:sub-tree}). Recall that $P^\star(h)$ is determined self-consistently from~\eref{P*}. 
Evidently, if $s_X$ is already up, the avalanche cannot propagate along the subtree rooted at $X$. 
It is straightforward to see that the probability $Q_0(h)$ that no avalanche propagates along a particular subtree --- i.e.,  the spin $s_X$ remains down even after its parent spin $s_Y$ flips up --- is given by
\begin{equation}
   Q_{0}(h) = \sum_{m=0}^{z-1} {{z-1} \choose m} \,[P^\star(h)]^m\, [1-P^\star(h)]^{\,z-1-m}\, [1-p_{m+1}(h)] .
   \label{Q_0}
\end{equation}

Now, for the root spin at $O$ to remain down at field $h$, but flip up when the field is increased to $h+dh$ (thereby initiating an avalanche), its local random field $h_O$ must satisfy  
$(z-2m)J - (h+dh) < h_O < (z-2m)J - h$,
which occurs with probability $\phi\!\left(zJ - 2mJ - h\right)\, dh$. 
Therefore, the generating function of the avalanche size distribution, 
\begin{equation}
   G(x,h) = \sum_{s=1}^{\infty} G_s(h)\, x^s ,
   \label{eq:Gx}
\end{equation}
can be expressed in terms of $Q(x,h)$ as
\begin{equation}
   G(x,h) = x \sum_{m=0}^{z} { z \choose m } \,[P^\star(h)]^m \,[Q(x,h)]^{\,z-m}\,\phi(zJ - 2mJ - h) .
   \label{G(x|h)}
\end{equation}

To summarize, for a given distribution $\phi(h_i)$ of the random fields, it is convenient to characterize the avalanche size distribution $G_s(h)$ through its generating function $G(x,h)$ defined in~\eref{eq:Gx}. The generating function is given explicitly by~\eref{G(x|h)}, where $P^\star(h)$ and $Q(x,h)$ are determined self-consistently from Eqs.~\eqref{P*} and~\eqref{Q(x)}, respectively. Finally, the coefficients $G_s(h)$ can, in principle, be obtained by inverting the generating function in the complex plane using Cauchy’s integral formula,
\begin{equation}
    G_s(h) = \frac{1}{2\pi i}\oint \frac{G(z,h)}{z^{s+1}}\, dz .
\end{equation}
In practice, it is, however, difficult to obtain $G_s(h)$ explicitly for general $\phi(h_i)$. Explicit calculations are possible for specific cases. For example, for a uniform distribution of random fields within the bounded supports $h_i \in [-\Delta, \Delta]$, for both coordination numbers  $z=2, 3$, the avalanche distribution $G_s(h)$ can be obtained explicitly for any $s$.

 For weak disorder, the system exhibits a macroscopic reversal, where the magnetization switches abruptly from $-1$ to $+1$ at a certain field value. In contrast, at higher disorder strengths, the response is mediated entirely through finite avalanches. In this regime, the magnetization curve is continuous, with no macroscopic jumps, and the avalanche-size distribution $G_s(h)$ decays exponentially for large $s$. 

For a generic continuous unimodal distribution of random fields with unbounded support, the system exhibits a first-order jump in magnetization as a function of the applied field when the coordination number satisfies $z \geq 4$ and the disorder strength is weak. In particular, for $z=4$, close to the discontinuity, i.e., as $[h_{\mathrm{disc}}-h]\to 0$, the avalanche-size distribution takes the asymptotic form  
\begin{equation}
   G_s(h) \sim s^{-3/2}\,\exp\!\left[-a\,(h_{\mathrm{disc}}-h)\,s\right], 
   \qquad \text{for large $s$},
\end{equation}
where $a$ is a constant. Exactly at the discontinuity $h= h_\mathrm{disc}$, the exponential cutoff disappears and the distribution exhibits a pure power law tail, 
\begin{equation}
   G_s(h_{\mathrm{disc}}) \sim s^{-3/2}.
\end{equation}

In numerical simulations, to reduce the statistical fluctuations, it is better to  calculate the distribution of avalanche sizes, averaged over the cycle.   It is easy to see that one gets a power law dependence $s^{-5/2}$ for the distribution of sizes of avalanches in the cycle-averaged data. Thus, we see that a robust power-law behavior  not requiring any fine-tuning, that is  the  defining feature of self-organized criticality,  occurs for the jump-size distribution in the Barkhausen noise, but  only for the cycle-averaged distribution, and only when disorder is weak enough and the hysteresis loop shows a macroscopic jump in magnetization.

\section{Concluding remarks}
\label{s:conclusion}

 In summary, we have provided a pedagogical review of hysteresis in magnetic systems. We  discussed two important early theoretical frameworks for understanding hysteresis: the Landau--Lifshitz--Gilbert (LLG) equation, which describes relaxation dynamics in ferromagnetic materials, and  the Preisach model, which treats a magnetic material as a composition of independent domains, each characterized by its own two-state magnetization and coercive field, whose random values are chosen independently. We  examined the various shapes of hysteresis loops exhibited by different types of systems, and the dependence of the loop area $A$ on the driving frequency $\omega$ and field amplitude $h_0$, expressed as $A \sim h_0^{\alpha} \omega^{\beta}$, where $\alpha$ and $\beta$ are the so-called \emph{Steinmetz coefficients}. We discussed simple cases where these coefficients can be determined exactly. Further, we reviewed hysteretic responses in Ising-like models, droplet-nucleation dynamics, and continuous spin models. In particular, we highlighted the dynamical phase transition of the cycle-averaged magnetization from zero to a non-zero value in the Ising case, and the transition from longitudinal to transverse magnetic ordering with increasing driving frequency in continuous spin models. Finally, we discussed hysteresis and Barkhausen noise in the zero-temperature random field Ising model, and the resulting distribution of avalanche (jump) sizes.

One can study hysteresis in other problems, for example, the difference response of a system in the heating and cooling cycles, when the temperature is varied linearly with time.  Fan and  Jinxiu discussed the thermal hysteresis and showed that the formalism developed in   \cite{ rao-krishnamurthy-pandit} can be adapted to this problem. In the model described by Ginzburg--Landau theory of an order parameter field $M$, and studied the mean-field case without noise \cite{fan-jinxiu}.  Here the system at any value of the control parameters tends to local minima of free energy using gradient descent, The analysis  turns out to quite similar to that in \cite{jung}, and they found that the area of the hysteresis loop $A$ in the temperature--entropy plane tends to a non-zero value, as R, the rate of temperature variation with time, tends to zero, as 
\begin{equation}
A = A_0 + K  |R|^{2/3}, {\rm ~~for ~~ small~~} R.
\end{equation}
The fact that $A_0$ does not tend to zero for $R$ tending to zero indicates the shortcoming of the assumed model, where the role of thermal noise in the evolution has been ignored, even though in this case,  the temperature is  varied.

Our discussion above can be easily extended to study the force--extension curves of a DNA chain subjected to periodic driving~\cite{dna1, dna2}. Stochastic resonance in a hysteretic system has been studied in~\cite{sides_stochastic, gade, acharyya_stochastic}.
One can also study hysteresis in quantum-mechanical spin models.  The simplest prototype is the one-dimensional Ising model subject to an oscillating time-dependent external field.  In this integrable model, the Hilbert space breakup into the 
 product of $(L/2)$ spaces of dimension $4$ each
 still holds when the field is time dependent, and the problem can be solved exactly~\cite{ref:transverseIsing}. 
Recently, there has been a lot of interest in periodically driven finite spin systems, in the context of  the quantum computers, and one may expect that studies of  hysteresis as discussed here  could act as stepping stones for the description  of the time-dependent Floquet states in the latter~\cite{floquet}. 
These are outside the scope of our discussion here.

\bmhead{Acknowledgements}
DD's work was supported by the  grant SP/DP/2023/658  by  Indian National Science Academy. We thank Rajaram Nityananda for critical comments on the draft manuscript, and Diptiman Sen for providing us with some relevant references.

\input{revisexx.bbl}

\end{document}

%% file: revisexx.bbl

%% file: revisexx.bbl
\begin{thebibliography}{61}
\ifx \bisbn   \undefined \def \bisbn  #1{ISBN #1}\fi
\ifx \binits  \undefined \def \binits#1{#1}\fi
\ifx \bauthor  \undefined \def \bauthor#1{#1}\fi
\ifx \batitle  \undefined \def \batitle#1{#1}\fi
\ifx \bjtitle  \undefined \def \bjtitle#1{#1}\fi
\ifx \bvolume  \undefined \def \bvolume#1{\textbf{#1}}\fi
\ifx \byear  \undefined \def \byear#1{#1}\fi
\ifx \bissue  \undefined \def \bissue#1{#1}\fi
\ifx \bfpage  \undefined \def \bfpage#1{#1}\fi
\ifx \blpage  \undefined \def \blpage #1{#1}\fi
\ifx \burl  \undefined \def \burl#1{\textsf{#1}}\fi
\ifx \doiurl  \undefined \def \doiurl#1{\url{https://doi.org/#1}}\fi
\ifx \betal  \undefined \def \betal{\textit{et al.}}\fi
\ifx \binstitute  \undefined \def \binstitute#1{#1}\fi
\ifx \binstitutionaled  \undefined \def \binstitutionaled#1{#1}\fi
\ifx \bctitle  \undefined \def \bctitle#1{#1}\fi
\ifx \beditor  \undefined \def \beditor#1{#1}\fi
\ifx \bpublisher  \undefined \def \bpublisher#1{#1}\fi
\ifx \bbtitle  \undefined \def \bbtitle#1{#1}\fi
\ifx \bedition  \undefined \def \bedition#1{#1}\fi
\ifx \bseriesno  \undefined \def \bseriesno#1{#1}\fi
\ifx \blocation  \undefined \def \blocation#1{#1}\fi
\ifx \bsertitle  \undefined \def \bsertitle#1{#1}\fi
\ifx \bsnm \undefined \def \bsnm#1{#1}\fi
\ifx \bsuffix \undefined \def \bsuffix#1{#1}\fi
\ifx \bparticle \undefined \def \bparticle#1{#1}\fi
\ifx \barticle \undefined \def \barticle#1{#1}\fi
\bibcommenthead
\ifx \bconfdate \undefined \def \bconfdate #1{#1}\fi
\ifx \botherref \undefined \def \botherref #1{#1}\fi
\ifx \url \undefined \def \url#1{\textsf{#1}}\fi
\ifx \bchapter \undefined \def \bchapter#1{#1}\fi
\ifx \bbook \undefined \def \bbook#1{#1}\fi
\ifx \bcomment \undefined \def \bcomment#1{#1}\fi
\ifx \oauthor \undefined \def \oauthor#1{#1}\fi
\ifx \citeauthoryear \undefined \def \citeauthoryear#1{#1}\fi
\ifx \endbibitem  \undefined \def \endbibitem {}\fi
\ifx \bconflocation  \undefined \def \bconflocation#1{#1}\fi
\ifx \arxivurl  \undefined \def \arxivurl#1{\textsf{#1}}\fi
\csname PreBibitemsHook\endcsname

\bibitem[\protect\citeauthoryear{Acharyya}{2005}]{review1}
\begin{barticle}
\bauthor{\bsnm{Acharyya}, \binits{M.}}:
\batitle{Nonequilibrium phase transitions in model ferromagnets: a review}.
\bjtitle{Int. J. Mod. Phys. C}
\bvolume{16},
\bfpage{1631}
(\byear{2005})
\doiurl{10.1142/S0129183105008266}
\end{barticle}
\endbibitem

\bibitem[\protect\citeauthoryear{Chakrabarti and Acharyya}{1999}]{review2}
\begin{barticle}
\bauthor{\bsnm{Chakrabarti}, \binits{B.K.}},
\bauthor{\bsnm{Acharyya}, \binits{M.}}:
\batitle{Dynamic transitions and hysteresis}.
\bjtitle{Rev. Mod. Phys.}
\bvolume{71},
\bfpage{847}
(\byear{1999})
\doiurl{10.1103/RevModPhys.71.847}
\end{barticle}
\endbibitem

\bibitem[\protect\citeauthoryear{Y\"uksel and Vatansever}{2021}]{review3}
\begin{barticle}
\bauthor{\bsnm{Y\"uksel}, \binits{Y.}},
\bauthor{\bsnm{Vatansever}, \binits{E.}}:
\batitle{Dynamic phase transition in classical {Ising} models}.
\bjtitle{J. Phys. D: Appl. Phys}
\bvolume{55},
\bfpage{073002}
(\byear{2021})
\doiurl{10.1088/1361-6463/ac2f6c}
\end{barticle}
\endbibitem

\bibitem[\protect\citeauthoryear{Lyuksyutov et~al.}{1999}]{lyuksyutov}
\begin{barticle}
\bauthor{\bsnm{Lyuksyutov}, \binits{I.F.}},
\bauthor{\bsnm{Nattermann}, \binits{T.}},
\bauthor{\bsnm{Pokrovsky}, \binits{V.}}:
\batitle{Theory of the hysteresis loop in ferromagnets}.
\bjtitle{Phys. Rev. B}
\bvolume{59},
\bfpage{4260}
(\byear{1999})
\doiurl{10.1103/PhysRevB.59.4260}
\end{barticle}
\endbibitem

\bibitem[\protect\citeauthoryear{Sides et~al.}{1998a}]{sides1}
\begin{barticle}
\bauthor{\bsnm{Sides}, \binits{S.W.}},
\bauthor{\bsnm{Rikvold}, \binits{P.A.}},
\bauthor{\bsnm{Novotny}, \binits{M.A.}}:
\batitle{Stochastic hysteresis and resonance in a kinetic {Ising} system}.
\bjtitle{Phys. Rev. E}
\bvolume{57},
\bfpage{6512}
(\byear{1998})
\doiurl{10.1103/PhysRevE.57.6512}
\end{barticle}
\endbibitem

\bibitem[\protect\citeauthoryear{Sides et~al.}{1998b}]{sides2a}
\begin{barticle}
\bauthor{\bsnm{Sides}, \binits{S.W.}},
\bauthor{\bsnm{Rikvold}, \binits{P.A.}},
\bauthor{\bsnm{Novotny}, \binits{M.A.}}:
\batitle{Kinetic {Ising} model in an oscillating field: Finite-size scaling at
  the dynamic phase transition}.
\bjtitle{Phys. Rev. Lett.}
\bvolume{81},
\bfpage{834}
(\byear{1998})
\doiurl{10.1103/PhysRevLett.81.834}
\end{barticle}
\endbibitem

\bibitem[\protect\citeauthoryear{Rikvold et~al.}{1994}]{sides2b}
\begin{barticle}
\bauthor{\bsnm{Rikvold}, \binits{P.A.}},
\bauthor{\bsnm{Tomita}, \binits{H.}},
\bauthor{\bsnm{Miyashita}, \binits{S.}},
\bauthor{\bsnm{Sides}, \binits{S.W.}}:
\batitle{Metastable lifetimes in a kinetic {Ising} model: Dependence on field
  and system size}.
\bjtitle{Phys. Rev. E}
\bvolume{49},
\bfpage{5080}
(\byear{1994})
\doiurl{10.1103/PhysRevE.49.5080}
\end{barticle}
\endbibitem

\bibitem[\protect\citeauthoryear{Sides et~al.}{1998}]{sides2c}
\begin{barticle}
\bauthor{\bsnm{Sides}, \binits{S.W.}},
\bauthor{\bsnm{Rikvold}, \binits{P.A.}},
\bauthor{\bsnm{Novotny}, \binits{M.A.}}:
\batitle{Hysteresis loop areas in kinetic {Ising} models: Effects of the
  switching mechanism}.
\bjtitle{J. Appl. Phys.}
\bvolume{83}(\bissue{11}),
\bfpage{6494}
(\byear{1998})
\doiurl{10.1063/1.367600}
\end{barticle}
\endbibitem

\bibitem[\protect\citeauthoryear{Colaiori}{2008}]{Colaiori08}
\begin{barticle}
\bauthor{\bsnm{Colaiori}, \binits{F.}}:
\batitle{Exactly solvable model of avalanches dynamics for {Barkhausen}
  crackling noise}.
\bjtitle{Advances in Physics}
\bvolume{57}(\bissue{4}),
\bfpage{287}
(\byear{2008})
\doiurl{10.1080/00018730802420614}
\end{barticle}
\endbibitem

\bibitem[\protect\citeauthoryear{Bertotti and
  Mayergoyz}{2006}]{mayergoyz2006science}
\begin{bbook}
\beditor{\bsnm{Bertotti}, \binits{G.}},
\beditor{\bsnm{Mayergoyz}, \binits{I.D.}} (eds.):
\bbtitle{The Science of Hysteresis}.
\bpublisher{Academic Press},
\blocation{Oxford}
(\byear{2006}).
\bcomment{Three volume set}.
\burl{https://www.sciencedirect.com/book/9780124808744/the-science-of-hysteresis}
\end{bbook}
\endbibitem

\bibitem[\protect\citeauthoryear{Skrotski{\u\i}}{1984}]{landau}
\begin{barticle}
\bauthor{\bsnm{Skrotski{\u\i}}, \binits{G.V.}}:
\batitle{The {Landau-Lifshitz} equation revisited}.
\bjtitle{Sov. Phys. Usp}
\bvolume{27},
\bfpage{977}
(\byear{1984})
\doiurl{10.1070/PU1984v027n12ABEH004101}
\end{barticle}
\endbibitem

\bibitem[\protect\citeauthoryear{Lakshmanan}{2011}]{lakshmanan}
\begin{barticle}
\bauthor{\bsnm{Lakshmanan}, \binits{M.}}:
\batitle{The fascinating world of the {Landau-Lifshitz–Gilbert} equation: an
  overview}.
\bjtitle{Phil. Trans. R. Soc. A.}
\bvolume{369},
\bfpage{1280}
(\byear{2011})
\doiurl{10.1098/rsta.2010.0319}
\end{barticle}
\endbibitem

\bibitem[\protect\citeauthoryear{Preisach}{1935}]{preisach}
\begin{barticle}
\bauthor{\bsnm{Preisach}, \binits{F.}}:
\batitle{{\"U}ber die magnetische nachwirkung}.
\bjtitle{Z. Physik}
\bvolume{94},
\bfpage{277}
(\byear{1935})
\doiurl{10.1007/BF01349418}
\end{barticle}
\endbibitem

\bibitem[\protect\citeauthoryear{Visintin}{2013}]{vistitin}
\begin{bbook}
\bauthor{\bsnm{Visintin}, \binits{A.}}:
\bbtitle{Differential Models of Hysteresis}
vol. \bseriesno{111}.
\bpublisher{Springer},
\blocation{Berlin and Heidelberg}
(\byear{2013}).
\doiurl{10.1007/978-3-662-11557-2}
\end{bbook}
\endbibitem

\bibitem[\protect\citeauthoryear{Mayergoyz}{2003}]{mayergoyz}
\begin{bbook}
\bauthor{\bsnm{Mayergoyz}, \binits{I.D.}}:
\bbtitle{Mathematical Models of Hysteresis and Their Applications},
\bedition{2}nd edn.
\bpublisher{Academic Press},
\blocation{San Diego}
(\byear{2003}).
\doiurl{10.1016/B978-0-12-480873-7.X5000-2}
\end{bbook}
\endbibitem

\bibitem[\protect\citeauthoryear{Steinmetz}{1984}]{steinmetz}
\begin{barticle}
\bauthor{\bsnm{Steinmetz}, \binits{C.P.}}:
\batitle{On the law of hysteresis}.
\bjtitle{Proc. IEEE}
\bvolume{72},
\bfpage{197}
(\byear{1984})
\doiurl{10.1109/PROC.1984.12842}
\end{barticle}
\endbibitem

\bibitem[\protect\citeauthoryear{Acharyya and Chakrabarti}{1995}]{acharyya2}
\begin{barticle}
\bauthor{\bsnm{Acharyya}, \binits{M.}},
\bauthor{\bsnm{Chakrabarti}, \binits{B.K.}}:
\batitle{Response of {Ising} systems to oscillating and pulsed fields:
  Hysteresis, ac, and pulse susceptibility}.
\bjtitle{Phys. Rev. B}
\bvolume{52},
\bfpage{6550}
(\byear{1995})
\doiurl{10.1103/PhysRevB.52.6550}
\end{barticle}
\endbibitem

\bibitem[\protect\citeauthoryear{Kim and Kim}{1997}]{kim}
\begin{barticle}
\bauthor{\bsnm{Kim}, \binits{Y.-H.}},
\bauthor{\bsnm{Kim}, \binits{J.-J.}}:
\batitle{Scaling behavior of an antiferroelectric hysteresis loop}.
\bjtitle{Phys. Rev. B}
\bvolume{55},
\bfpage{11933}
(\byear{1997})
\doiurl{10.1103/PhysRevB.55.R11933}
\end{barticle}
\endbibitem

\bibitem[\protect\citeauthoryear{Glauber}{1963}]{Glauber1963}
\begin{barticle}
\bauthor{\bsnm{Glauber}, \binits{R.J.}}:
\batitle{Time‐dependent statistics of the {I}sing model}.
\bjtitle{J. Math. Phys.}
\bvolume{4},
\bfpage{294}
(\byear{1963})
\doiurl{10.1063/1.1703954}
\end{barticle}
\endbibitem

\bibitem[\protect\citeauthoryear{Jung et~al.}{1990}]{jung}
\begin{barticle}
\bauthor{\bsnm{Jung}, \binits{P.}},
\bauthor{\bsnm{Gray}, \binits{G.}},
\bauthor{\bsnm{Roy}, \binits{R.}},
\bauthor{\bsnm{Mandel}, \binits{P.}}:
\batitle{Scaling law for dynamical hysteresis}.
\bjtitle{Phys. Rev. Lett.}
\bvolume{65},
\bfpage{1873}
(\byear{1990})
\doiurl{10.1103/PhysRevLett.65.1873}
\end{barticle}
\endbibitem

\bibitem[\protect\citeauthoryear{Broner et~al.}{1997}]{goldzstein}
\begin{barticle}
\bauthor{\bsnm{Broner}, \binits{F.}},
\bauthor{\bsnm{Goldsztein}, \binits{G.H.}},
\bauthor{\bsnm{Strogatz}, \binits{S.H.}}:
\batitle{Dynamical hysteresis without static hysteresis: Scaling laws and
  asymptotic expansions}.
\bjtitle{SIAM J. Appl. Math.}
\bvolume{57},
\bfpage{1163}
(\byear{1997})
\doiurl{10.1137/S0036139995290733}
{\href{https://arxiv.org/abs/https://doi.org/10.1137/S0036139995290733}{{https://doi.org/10.1137/S0036139995290733}}}
\end{barticle}
\endbibitem

\bibitem[\protect\citeauthoryear{Oxtoby}{1992}]{nucleation}
\begin{barticle}
\bauthor{\bsnm{Oxtoby}, \binits{D.W.}}:
\batitle{Homogeneous nucleation: theory and experiment}.
\bjtitle{J. Phys.: Condens. Matter}
\bvolume{4}(\bissue{38}),
\bfpage{7627}
(\byear{1992})
\doiurl{10.1088/0953-8984/4/38/001}
\end{barticle}
\endbibitem

\bibitem[\protect\citeauthoryear{Thomas and Dhar}{1993}]{dhar:93}
\begin{barticle}
\bauthor{\bsnm{Thomas}, \binits{P.B.}},
\bauthor{\bsnm{Dhar}, \binits{D.}}:
\batitle{Hysteresis in isotropic spin systems}.
\bjtitle{J. Phys. A: Math. Gen.}
\bvolume{26},
\bfpage{3973}
(\byear{1993})
\doiurl{10.1088/0305-4470/26/16/014}
\end{barticle}
\endbibitem

\bibitem[\protect\citeauthoryear{He and Wang}{1993}]{he-wang}
\begin{barticle}
\bauthor{\bsnm{He}, \binits{Y.-L.}},
\bauthor{\bsnm{Wang}, \binits{G.-C.}}:
\batitle{Observation of dynamic scaling of magnetic hysteresis in ultrathin
  ferromagnetic fe/au(001) films}.
\bjtitle{Phys. Rev. Lett.}
\bvolume{70},
\bfpage{2336}
(\byear{1993})
\doiurl{10.1103/PhysRevLett.70.2336}
\end{barticle}
\endbibitem

\bibitem[\protect\citeauthoryear{Jiang et~al.}{1995}]{jiang-yang-wang}
\begin{barticle}
\bauthor{\bsnm{Jiang}, \binits{Q.}},
\bauthor{\bsnm{Yang}, \binits{H.-N.}},
\bauthor{\bsnm{Wang}, \binits{G.-C.}}:
\batitle{Scaling and dynamics of low-frequency hysteresis loops in ultrathin co
  films on a cu(001) surface}.
\bjtitle{Phys. Rev. B}
\bvolume{52},
\bfpage{14911}
(\byear{1995})
\doiurl{10.1103/PhysRevB.52.14911}
\end{barticle}
\endbibitem

\bibitem[\protect\citeauthoryear{Suen and Erskine}{1997}]{suen-erskine}
\begin{barticle}
\bauthor{\bsnm{Suen}, \binits{J.-S.}},
\bauthor{\bsnm{Erskine}, \binits{J.L.}}:
\batitle{Magnetic hysteresis dynamics: Thin
  $\mathit{p}(1\ifmmode\times\else\texttimes\fi{}1)$ fe films on flat and
  stepped w(110)}.
\bjtitle{Phys. Rev. Lett.}
\bvolume{78},
\bfpage{3567}
(\byear{1997})
\doiurl{10.1103/PhysRevLett.78.3567}
\end{barticle}
\endbibitem

\bibitem[\protect\citeauthoryear{Sengupta et~al.}{1992}]{sengupta92}
\begin{barticle}
\bauthor{\bsnm{Sengupta}, \binits{S.}},
\bauthor{\bsnm{Marathe}, \binits{Y.}},
\bauthor{\bsnm{Puri}, \binits{S.}}:
\batitle{Cell-dynamical simulation of magnetic hysteresis in the
  two-dimensional ising system}.
\bjtitle{Phys. Rev. B}
\bvolume{45},
\bfpage{7828}
(\byear{1992})
\doiurl{10.1103/PhysRevB.45.7828}
\end{barticle}
\endbibitem

\bibitem[\protect\citeauthoryear{Rao et~al.}{1990}]{rao-krishnamurthy-pandit}
\begin{barticle}
\bauthor{\bsnm{Rao}, \binits{M.}},
\bauthor{\bsnm{Krishnamurthy}, \binits{H.R.}},
\bauthor{\bsnm{Pandit}, \binits{R.}}:
\batitle{Magnetic hysteresis in two model spin systems}.
\bjtitle{Phys. Rev. B}
\bvolume{42},
\bfpage{856}
(\byear{1990})
\doiurl{10.1103/PhysRevB.42.856}
\end{barticle}
\endbibitem

\bibitem[\protect\citeauthoryear{Dhar and Thomas}{1992}]{dhar:92}
\begin{barticle}
\bauthor{\bsnm{Dhar}, \binits{D.}},
\bauthor{\bsnm{Thomas}, \binits{P.B.}}:
\batitle{{Hysteresis and self-organized criticality in the O(N) model in the
  limit N to infinity}}.
\bjtitle{J. Phys. A: Math. Gen.}
\bvolume{25},
\bfpage{4967}
(\byear{1992})
\doiurl{10.1088/0305-4470/25/19/012}
\end{barticle}
\endbibitem

\bibitem[\protect\citeauthoryear{Somoza and Desai}{1993}]{somoza:93}
\begin{barticle}
\bauthor{\bsnm{Somoza}, \binits{A.M.}},
\bauthor{\bsnm{Desai}, \binits{R.C.}}:
\batitle{Kinetics of systems with continuous symmetry under the effect of an
  external field}.
\bjtitle{Phys. Rev. Lett.}
\bvolume{70},
\bfpage{3279}
(\byear{1993})
\doiurl{10.1103/PhysRevLett.70.3279}
\end{barticle}
\endbibitem

\bibitem[\protect\citeauthoryear{Mahato and Shenoy}{1993}]{mahato}
\begin{barticle}
\bauthor{\bsnm{Mahato}, \binits{M.C.}},
\bauthor{\bsnm{Shenoy}, \binits{S.R.}}:
\batitle{Langevin dynamic simulation of hysteresis in a field-swept {L}andau
  potential}.
\bjtitle{J Stat. Phys.}
\bvolume{73},
\bfpage{123}
(\byear{1993})
\doiurl{10.1007/BF01052753}
\end{barticle}
\endbibitem

\bibitem[\protect\citeauthoryear{Scopa and Wald}{2018}]{scopa}
\begin{botherref}
\oauthor{\bsnm{Scopa}, \binits{S.}},
\oauthor{\bsnm{Wald}, \binits{S.}}:
Dynamical off-equilibrium scaling across magnetic first-order phase
  transitions.
J. Stat. Mech.,
113205
(2018)
\doiurl{10.1088/1742-5468/aaeb46}
\end{botherref}
\endbibitem

\bibitem[\protect\citeauthoryear{Dutta}{2004}]{dutta}
\begin{barticle}
\bauthor{\bsnm{Dutta}, \binits{S.B.}}:
\batitle{Phase transitions in periodically driven macroscopic systems}.
\bjtitle{Phys. Rev. E}
\bvolume{69},
\bfpage{066115}
(\byear{2004})
\doiurl{10.1103/PhysRevE.69.066115}
\end{barticle}
\endbibitem

\bibitem[\protect\citeauthoryear{Tom\'e and de~Oliveira}{1990}]{tome}
\begin{barticle}
\bauthor{\bsnm{Tom\'e}, \binits{T.}},
\bauthor{\bsnm{Oliveira}, \binits{M.J.}}:
\batitle{Dynamic phase transition in the kinetic {Ising} model under a
  time-dependent oscillating field}.
\bjtitle{Phys. Rev. A}
\bvolume{41},
\bfpage{4251}
(\byear{1990})
\doiurl{10.1103/PhysRevA.41.4251}
\end{barticle}
\endbibitem

\bibitem[\protect\citeauthoryear{Junier and Kurchan}{2003}]{junier}
\begin{barticle}
\bauthor{\bsnm{Junier}, \binits{I.}},
\bauthor{\bsnm{Kurchan}, \binits{J.}}:
\batitle{Tailoring symmetry groups using external alternate fields}.
\bjtitle{Europhys. Lett.}
\bvolume{63},
\bfpage{674}
(\byear{2003})
\doiurl{10.1209/epl/i2003-00583-2}
\end{barticle}
\endbibitem

\bibitem[\protect\citeauthoryear{Paessens and Henkel}{2003}]{paessens}
\begin{barticle}
\bauthor{\bsnm{Paessens}, \binits{M.}},
\bauthor{\bsnm{Henkel}, \binits{M.}}:
\batitle{The kinetic spherical model in a magnetic field}.
\bjtitle{J. Phys. A: Math. Gen.}
\bvolume{36},
\bfpage{8983}
(\byear{2003})
\doiurl{10.1088/0305-4470/36/34/304}
\end{barticle}
\endbibitem

\bibitem[\protect\citeauthoryear{Barkhausen}{1919}]{barkhausen:19}
\begin{barticle}
\bauthor{\bsnm{Barkhausen}, \binits{H.}}:
\batitle{Zwei mit hilfe der neuen verst{\"a}rker entdeckte erscheinungen}.
\bjtitle{Phys. Z}
\bvolume{20},
\bfpage{401}
(\byear{1919})
\end{barticle}
\endbibitem

\bibitem[\protect\citeauthoryear{Feynman et~al.}{1977}]{feynman:77}
\begin{bbook}
\bauthor{\bsnm{Feynman}, \binits{R.P.}},
\bauthor{\bsnm{Leighton}, \binits{R.B.}},
\bauthor{\bsnm{Sands}, \binits{M.}}:
\bbtitle{The Feynman Lectures on Physics}
vol. \bseriesno{II}.
\bpublisher{Addison-Wesley},
\blocation{{}}
(\byear{1977})
\end{bbook}
\endbibitem

\bibitem[\protect\citeauthoryear{Sipahi}{1994}]{sipahi:94}
\begin{barticle}
\bauthor{\bsnm{Sipahi}, \binits{L.B.}}:
\batitle{Overview of applications of micromagnetic barkhausen emissions as
  noninvasive material characterization technique}.
\bjtitle{J. Appl. Phys.}
\bvolume{75}(\bissue{10}),
\bfpage{6978}
(\byear{1994})
\doiurl{10.1063/1.356747}
\end{barticle}
\endbibitem

\bibitem[\protect\citeauthoryear{Spasojevi\'c et~al.}{1996}]{spasojevic:96}
\begin{barticle}
\bauthor{\bsnm{Spasojevi\'c}, \binits{D.}},
\bauthor{\bsnm{Bukvi\'c}, \binits{S.}},
\bauthor{\bsnm{Milo\v{s}evi\'c}, \binits{S.}},
\bauthor{\bsnm{Stanley}, \binits{H.E.}}:
\batitle{Barkhausen noise: Elementary signals, power laws, and scaling
  relations}.
\bjtitle{Phys. Rev. E}
\bvolume{54},
\bfpage{2531}
(\byear{1996})
\doiurl{10.1103/PhysRevE.54.2531}
\end{barticle}
\endbibitem

\bibitem[\protect\citeauthoryear{Sabhapandit}{2002}]{sabhapandit:phd}
\begin{botherref}
\oauthor{\bsnm{Sabhapandit}, \binits{S.}}:
Avalanches in driven systems.
PhD thesis,
Tata Institute of Fundamental Research,
Mumbai, India
(2002).
\doiurl{10.48550/arXiv.cond-mat/0209569} .
arXiv:cond-mat/0209569
\end{botherref}
\endbibitem

\bibitem[\protect\citeauthoryear{Urbach et~al.}{1995}]{urbach:95}
\begin{barticle}
\bauthor{\bsnm{Urbach}, \binits{J.S.}},
\bauthor{\bsnm{Madison}, \binits{R.C.}},
\bauthor{\bsnm{Markert}, \binits{J.T.}}:
\batitle{Interface depinning, self-organized criticality, and the barkhausen
  effect}.
\bjtitle{Phys. Rev. Lett.}
\bvolume{75},
\bfpage{276}
(\byear{1995})
\doiurl{10.1103/PhysRevLett.75.276}
\end{barticle}
\endbibitem

\bibitem[\protect\citeauthoryear{Imry and Ma}{1975}]{imry:75}
\begin{barticle}
\bauthor{\bsnm{Imry}, \binits{Y.}},
\bauthor{\bsnm{Ma}, \binits{S.-k.}}:
\batitle{Random-field instability of the ordered state of continuous symmetry}.
\bjtitle{Phys. Rev. Lett.}
\bvolume{35},
\bfpage{1399}
(\byear{1975})
\doiurl{10.1103/PhysRevLett.35.1399}
\end{barticle}
\endbibitem

\bibitem[\protect\citeauthoryear{Imbrie}{1984}]{imbrie:84}
\begin{barticle}
\bauthor{\bsnm{Imbrie}, \binits{J.Z.}}:
\batitle{Lower critical dimension of the random-field {Ising} model}.
\bjtitle{Phys. Rev. Lett.}
\bvolume{53},
\bfpage{1747}
(\byear{1984})
\doiurl{10.1103/PhysRevLett.53.1747}
\end{barticle}
\endbibitem

\bibitem[\protect\citeauthoryear{Aizenman and Wehr}{1989}]{aizenmann:89}
\begin{barticle}
\bauthor{\bsnm{Aizenman}, \binits{M.}},
\bauthor{\bsnm{Wehr}, \binits{J.}}:
\batitle{Rounding of first-order phase transitions in systems with quenched
  disorder}.
\bjtitle{Phys. Rev. Lett.}
\bvolume{62},
\bfpage{2503}
(\byear{1989})
\doiurl{10.1103/PhysRevLett.62.2503}
\end{barticle}
\endbibitem

\bibitem[\protect\citeauthoryear{Nattermann}{1998}]{nattermann1998theory}
\begin{bbook}
\bauthor{\bsnm{Nattermann}, \binits{T.}}:
\bbtitle{Theory of the random field {Ising} model},
pp. \bfpage{277}--\blpage{298}.
\bpublisher{World Scientific},
\blocation{{}}
(\byear{1998}).
\doiurl{10.1142/9789812819437_0009}
\end{bbook}
\endbibitem

\bibitem[\protect\citeauthoryear{Kurbah and Shukla}{2011}]{kurbah:2011}
\begin{barticle}
\bauthor{\bsnm{Kurbah}, \binits{L.}},
\bauthor{\bsnm{Shukla}, \binits{P.}}:
\batitle{Hysteresis in the antiferromagnetic random-field ising model at zero
  temperature}.
\bjtitle{Phys. Rev. E}
\bvolume{83},
\bfpage{061136}
(\byear{2011})
\doiurl{10.1103/PhysRevE.83.061136}
\end{barticle}
\endbibitem

\bibitem[\protect\citeauthoryear{Salvat-Pujol et~al.}{2009}]{salvat-pujol:2009}
\begin{barticle}
\bauthor{\bsnm{Salvat-Pujol}, \binits{F.}},
\bauthor{\bsnm{Vives}, \binits{E.}},
\bauthor{\bsnm{Rosinberg}, \binits{M.-L.}}:
\batitle{{Hysteresis in the $T=0$ random-field Ising model: Beyond metastable
  dynamics}}.
\bjtitle{Phys. Rev. E}
\bvolume{79},
\bfpage{061116}
(\byear{2009})
\doiurl{10.1103/PhysRevE.79.061116}
\end{barticle}
\endbibitem

\bibitem[\protect\citeauthoryear{Sethna et~al.}{1993}]{sethna:93}
\begin{barticle}
\bauthor{\bsnm{Sethna}, \binits{J.P.}},
\bauthor{\bsnm{Dahmen}, \binits{K.}},
\bauthor{\bsnm{Kartha}, \binits{S.}},
\bauthor{\bsnm{Krumhansl}, \binits{J.A.}},
\bauthor{\bsnm{Roberts}, \binits{B.W.}},
\bauthor{\bsnm{Shore}, \binits{J.D.}}:
\batitle{Hysteresis and hierarchies: Dynamics of disorder-driven first-order
  phase transformations}.
\bjtitle{Phys. Rev. Lett.}
\bvolume{70},
\bfpage{3347}
(\byear{1993})
\doiurl{10.1103/PhysRevLett.70.3347}
\end{barticle}
\endbibitem

\bibitem[\protect\citeauthoryear{Dhar et~al.}{1997}]{dhar:97}
\begin{barticle}
\bauthor{\bsnm{Dhar}, \binits{D.}},
\bauthor{\bsnm{Shukla}, \binits{P.}},
\bauthor{\bsnm{Sethna}, \binits{J.P.}}:
\batitle{Zero-temperature hysteresis in the random-field {Ising} model on a
  {Bethe} lattice}.
\bjtitle{J. Phys. A: Math. Gen.}
\bvolume{30},
\bfpage{5259}
(\byear{1997})
\doiurl{10.1088/0305-4470/30/15/013}
\end{barticle}
\endbibitem

\bibitem[\protect\citeauthoryear{Shukla}{2000}]{shukla:00}
\begin{barticle}
\bauthor{\bsnm{Shukla}, \binits{P.}}:
\batitle{Exact solution of return hysteresis loops in a one-dimensional
  random-field {Ising} model at zero temperature}.
\bjtitle{Phys. Rev. E}
\bvolume{62},
\bfpage{4725}
(\byear{2000})
\doiurl{10.1103/PhysRevE.62.4725}
\end{barticle}
\endbibitem

\bibitem[\protect\citeauthoryear{Shukla}{2001}]{shukla:01}
\begin{barticle}
\bauthor{\bsnm{Shukla}, \binits{P.}}:
\batitle{Exact expressions for minor hysteresis loops in the random field
  {Ising} model on a {Bethe} lattice at zero temperature}.
\bjtitle{Phys. Rev. E}
\bvolume{63},
\bfpage{027102}
(\byear{2001})
\doiurl{10.1103/PhysRevE.63.027102}
\end{barticle}
\endbibitem

\bibitem[\protect\citeauthoryear{Sabhapandit et~al.}{2000}]{sabhapandit:00}
\begin{barticle}
\bauthor{\bsnm{Sabhapandit}, \binits{S.}},
\bauthor{\bsnm{Shukla}, \binits{P.}},
\bauthor{\bsnm{Dhar}, \binits{D.}}:
\batitle{Distribution of avalanche sizes in the hysteretic response of the
  random-field {Ising} model on a {Bethe} lattice at zero temperature}.
\bjtitle{J. Stat. Phys.}
\bvolume{98}(\bissue{1}),
\bfpage{103}
(\byear{2000})
\doiurl{10.1023/A:1018622805347}
\end{barticle}
\endbibitem

\bibitem[\protect\citeauthoryear{Fan and Jinxiu}{1995}]{fan-jinxiu}
\begin{barticle}
\bauthor{\bsnm{Fan}, \binits{Z.}},
\bauthor{\bsnm{Jinxiu}, \binits{Z.}}:
\batitle{Scaling of thermal hysteresis with temperature scanning rate}.
\bjtitle{Phys. Rev. E}
\bvolume{51},
\bfpage{2898}
(\byear{1995})
\doiurl{10.1103/PhysRevE.51.2898}
\end{barticle}
\endbibitem

\bibitem[\protect\citeauthoryear{Kapri}{2012}]{dna1}
\begin{barticle}
\bauthor{\bsnm{Kapri}, \binits{R.}}:
\batitle{Hysteresis and nonequilibrium work theorem for dna unzipping}.
\bjtitle{Phys. Rev. E}
\bvolume{86},
\bfpage{041906}
(\byear{2012})
\doiurl{10.1103/PhysRevE.86.041906}
\end{barticle}
\endbibitem

\bibitem[\protect\citeauthoryear{Kumar et~al.}{2016}]{dna2}
\begin{barticle}
\bauthor{\bsnm{Kumar}, \binits{S.}},
\bauthor{\bsnm{Kumar}, \binits{R.}},
\bauthor{\bsnm{Janke}, \binits{W.}}:
\batitle{Periodically driven dna: Theory and simulation}.
\bjtitle{Phys. Rev. E}
\bvolume{93},
\bfpage{010402}
(\byear{2016})
\doiurl{10.1103/PhysRevE.93.010402}
\end{barticle}
\endbibitem

\bibitem[\protect\citeauthoryear{Sides et~al.}{1998}]{sides_stochastic}
\begin{barticle}
\bauthor{\bsnm{Sides}, \binits{S.W.}},
\bauthor{\bsnm{Rikvold}, \binits{P.A.}},
\bauthor{\bsnm{Novotny}, \binits{M.A.}}:
\batitle{Stochastic hysteresis and resonance in a kinetic {Ising} system}.
\bjtitle{Phys. Rev. E}
\bvolume{57},
\bfpage{6512}
(\byear{1998})
\doiurl{10.1103/PhysRevE.57.6512}
\end{barticle}
\endbibitem

\bibitem[\protect\citeauthoryear{Gade et~al.}{1997}]{gade}
\begin{barticle}
\bauthor{\bsnm{Gade}, \binits{P.M.}},
\bauthor{\bsnm{Rai}, \binits{R.}},
\bauthor{\bsnm{Singh}, \binits{H.}}:
\batitle{Stochastic resonance in maps and coupled map lattices}.
\bjtitle{Phys. Rev. E}
\bvolume{56},
\bfpage{2518}
(\byear{1997})
\doiurl{10.1103/PhysRevE.56.2518}
\end{barticle}
\endbibitem

\bibitem[\protect\citeauthoryear{Acharyya}{1999}]{acharyya_stochastic}
\begin{barticle}
\bauthor{\bsnm{Acharyya}, \binits{M.}}:
\batitle{Nonequilibrium phase transition in the kinetic {Ising} model:
  Existence of a tricritical point and stochastic resonance}.
\bjtitle{Phys. Rev. E}
\bvolume{59},
\bfpage{218}
(\byear{1999})
\doiurl{10.1103/PhysRevE.59.218}
\end{barticle}
\endbibitem

\bibitem[\protect\citeauthoryear{Russomanno et~al.}{2012}]{ref:transverseIsing}
\begin{barticle}
\bauthor{\bsnm{Russomanno}, \binits{A.}},
\bauthor{\bsnm{Silva}, \binits{A.}},
\bauthor{\bsnm{Santoro}, \binits{G.E.}}:
\batitle{Periodic steady regime and interference in a periodically driven
  quantum system}.
\bjtitle{Phys. Rev. Lett.}
\bvolume{109},
\bfpage{257201}
(\byear{2012})
\doiurl{10.1103/PhysRevLett.109.257201}
\end{barticle}
\endbibitem

\bibitem[\protect\citeauthoryear{Eckardt}{2017}]{floquet}
\begin{barticle}
\bauthor{\bsnm{Eckardt}, \binits{A.}}:
\batitle{Colloquium: Atomic quantum gases in periodically driven optical
  lattices}.
\bjtitle{Rev. Mod. Phys.}
\bvolume{89},
\bfpage{011004}
(\byear{2017})
\doiurl{10.1103/RevModPhys.89.011004}
\end{barticle}
\endbibitem

\end{thebibliography}
